\documentclass[preprint,11pt]{elsarticle}

\usepackage{amssymb}
\usepackage{latexsym}
\usepackage{wrapfig}
\usepackage{amsmath}
\usepackage{booktabs}
\usepackage{threeparttable}
\usepackage{xcolor}
\usepackage{amsthm}
\usepackage{comment}
\usepackage{graphicx, caption, subcaption}
\usepackage{titlesec,enumitem}
\usepackage{tabularx}
\usepackage{array}
\usepackage{xcolor}
\usepackage{makecell}
\usepackage{placeins}
\usepackage{caption}
\usepackage{footnote}
\usepackage{hyperref}

\biboptions{numbers,sort&compress}
\allowdisplaybreaks

\newcommand{\beq}{\begin{equation}}
\newcommand{\eeq}{\end{equation}}
\newcommand{\beqs}{\begin{eqnarray}}
\newcommand{\eeqs}{\end{eqnarray}}

\newtheorem{theorem*}{Theorem}

\DeclareMathOperator{\sym}{sym}

\DeclareMathOperator{\Tr}{Tr}

\journal{\ldots}

\let\today\relax
\makeatletter
\def\ps@pprintTitle{%
    \let\@oddhead\@empty
    \let\@evenhead\@empty
    \def\@oddfoot{\footnotesize\itshape
         { } \hfill\today}%
    \let\@evenfoot\@oddfoot
    }
\makeatother

\hypersetup{
    colorlinks=true,
    linkcolor=black,
    citecolor=black,
    filecolor=black,
    urlcolor=black,
}

\begin{document}

\begin{frontmatter}

%% Title, authors and addresses

%% use the tnoteref command within \title for footnotes;
%% use the tnotetext command for theassociated footnote;
%% use the fnref command within \author or \address for footnotes;
%% use the fntext command for theassociated footnote;
%% use the corref command within \author for corresponding author footnotes;
%% use the cortext command for theassociated footnote;
%% use the ead command for the email address,
%% and the form \ead[url] for the home page:
%% \title{Title\tnoteref{label1}}
%% \tnotetext[label1]{}
%% \author{Name\corref{cor1}\fnref{label2}}
%% \ead{email address}
%% \ead[url]{home page}
%% \fntext[label2]{}
%% \cortext[cor1]{}
%% \address{Address\fnref{label3}}
%% \fntext[label3]{}

\title{Modelling planar kirigami metamaterials as generalized elastic continua}
%\title{Modeling planar kirigami metamaterials as a micromorphic elastic continuum}

%% use optional labels to link authors explicitly to addresses:
%% \author[label1,label2]{}
%% \address[label1]{}
%% \address[label2]{}

\author[1]{Yue Zheng\corref{mycorrespondingauthor}}

\author[2]{Imtiar Niloy}
\author[3]{Ian Tobasco}
\author[2]{Paolo Celli}
\author[1]{Paul Plucinsky\corref{mycorrespondingauthor}}
\ead{plucinsk@usc.edu}

\address[1]{Aerospace and Mechanical Engineering, University of Southern California, Los Angeles, California 90089, USA}
\address[2]{Civil Engineering, Stony Brook University, Stony Brook, NY 11794, USA}
\address[3]{Mathematics, Statistics, and Computer Science, University of Illinois at Chicago, Chicago, IL 60607, USA}

\begin{abstract}

Kirigami metamaterials dramatically change their shape through a coordinated motion of nearly rigid panels and flexible slits. Here, we study a model system for mechanism-based planar kirigami featuring periodic patterns of quadrilateral panels and rhombi slits, with the goal of predicting their engineering scale response to a broad range of loads. We develop a generalized continuum model based on the kirigami's effective (cell-averaged) nonlinear deformation, along with its slit actuation and gradients thereof. The model accounts for three sources of elasticity: a strong preference for the effective fields to match those of a local mechanism, inter-panel stresses arising from gradients in slit actuation, and  distributed hinge bending. We provide a finite element formulation of this model and implement it using the commercial software Abaqus. Simulations of the model agree {\color{black}quantitatively} with experiments across designs and loading conditions.
\end{abstract}

\begin{keyword}
%% keywords here, in the form: keyword \sep keyword
Metamaterials \sep Kirigami \sep Continuum modeling \sep FEM 
%% PACS codes here, in the form: \PACS code \sep code

%% MSC codes here, in the form: \MSC code \sep code
%% or \MSC[2008] code \sep code (2000 is the default)

\end{keyword}

\end{frontmatter}

\vspace{5px}
\textbf{This article may be downloaded for personal use only. Any other use requires prior permission of the authors and the Royal Society}. This article appeared in: \emph{Proceedings of the Royal Society A: Mathematical, Physical and Engineering Sciences} {\bf 479}(2272), 20220665 (2023) {and may be found at}: \url{https://doi.org/10.1098/rspa.2022.0665}

%% main text
\section{Introduction}\label{sec:s1}

Mechanical metamaterials are solids whose global response is dominated by geometry and topology, rather than material physics. A typical mechanical metamaterial, like the Miura-Origami \cite{schenk2013geometry} or the rotating-squares lattice \cite{grima2000auxetic}, is built from a pattern of repeatingunit  cells. Each cell is composed of  stiff and flexible elements, whose layout enables for a bulk shape-morphing response to stimuli.  The geometric rules linking design to deformation in such systems have captivated theorists \cite{lubensky2015phonons,bertoldi2017flexible,liuetalOrigamiStructure}. In parallel, the embrace of shape-morphing in modern engineering --- for the design of  stents~\cite{Kuribayashi2006, Velvaluri2021},  soft robotic grippers \cite{shintake2018soft, rafsanjani2019programming}, deployable space structures~\cite{miura1985method,Arya2017}, and the like ---  has drawn applied researchers to these systems in an effort to demonstrate new functionalities.  Thus, mechanical metamaterials sit at the intersection of engineering design and mechanics, where new paradigms are needed to realize their full potential.
 
\begin{figure}[h!]
\centering
\includegraphics[scale=1.0]{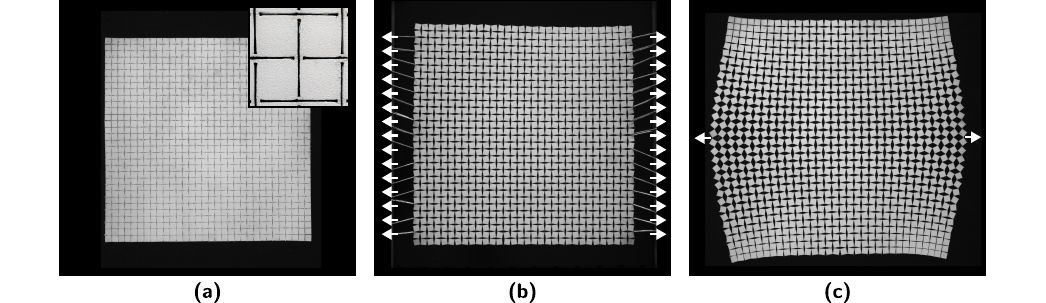}
\caption{(a) Rotating-squares mechanical metamaterial in its undeformed configuration. (b) Uniform response of the metamaterial to homogeneous loading. (c) Non-uniform, locally-mechanistic response to inhomogeneous loading.}
\label{fig:intro}
\end{figure}

This paper concerns one such paradigm, namely, how to best model the engineering scale response of a shape-morphing mechanical metamaterial under a broad range of loads. To illustrate the challenges, consider the example in Fig.\;\ref{fig:intro}. 
The pattern, a rotating-squares architecture, is composed of a repeating unit cell of four quadrilateral panels, shown in Fig.\;\ref{fig:intro}(a). This design exhibits a single \textit{mechanism} \cite{grima2000auxetic}, which counter-rotates the panels periodically and results in an effectively uniform overall motion in response to homogeneous loads (Fig.\;\ref{fig:intro}(b)). However, when subjected to inhomogeneous loads (Fig.\;\ref{fig:intro}(c)), the slits actuate in a non-uniform way --- one that is \textit{locally mechanistic} rather than globally so --- reflecting the interaction between the design of the unit cell and the choice of applied loads. 

Behind this response is a complex interplay between geometry and elasticity.  Two modeling  approaches  are common in the literature, the first of which is purely geometric: panels are taken to be rigid and connected by ideal hinges (folds, in origami), with the goal of characterizing rigid deformations and mechanisms using kinematic compatibility. This approach, first popularized in the origami literature \cite{huffman1976curvature,hull2002modelling,tachi2009generalization}, has rationalized the design of mechanical metamaterials and illuminated their basic mechanisms \cite{chen2016symmetric,lang2018rigidly,feng2020designs,dieleman2020jigsaw,dudte2021additive,choi2021compact,dang2022theorem,walker2022algorithmic}. However, it does not include elasticity. 
   
The second modeling approach is based on long-established structural mechanics ideas~\cite{pellegrino1986matrix, schenk2011origami} exemplified in recent literature by the bar-hinge method \cite{filipov2017bar,liu2017nonlinear} and related spring methods \cite{coulais2018characteristic, deng2020characterization}. This approach replaces the metamaterial with assemblies of bars and hinges, whose geometric arrangement encodes desired morphing attributes. Elasticity is accounted for by modeling bars/hinges as linear/torsional springs, and the structure's response is analyzed using standard numerical methods. While this approach is versatile and convenient for systems composed of a small number of building blocks, it can become computationally expensive for larger systems with many unit cells. Additionally, fitting spring stiffnesses to yield accurate global behaviors can be challenging.
   
Recently, a third approach has emerged from the idea of finding  effective, coarse-grained models for the deformations of  origami and kirigami   \cite{wei2013geometric, nassar2017curvature, khajehtourian2021continuum, czajkowski2022conformal, nassar2022strain, MCMAHAN2022, mcinerney2021discrete}.  In this approach, the metamaterial is modeled as an effectively continuum object, with the aim of capturing the collective response of its cells through averaging. Supporting this idea is the  heuristic that origami and kirigami have many \textit{soft modes}, as in Fig.\;\ref{fig:intro}(c), that resemble  mechanisms locally but describe global, non-mechanistic shape change. Focusing on kirigami, we highlight the works  \cite{czajkowski2022conformal,MCMAHAN2022,deng2020characterization}, which are closest to what we do here.     The works of Czajkowski \textit{et al.}\;and McMahan \textit{et al.}\;embrace the soft modes heuristic by introducing elastic models whose stresses drive the kirigami's effective deformations towards cell-averaged local mechanisms. These  models account for the geometric nonlinearity inherent to soft modes. However, they are based solely on deformations, and miss the fact that some sources of elastic frustration in kirigami arise instead from ``micro''-incompatibilities.

\begin{figure}[h!]
\centering
\includegraphics[scale=1.0]{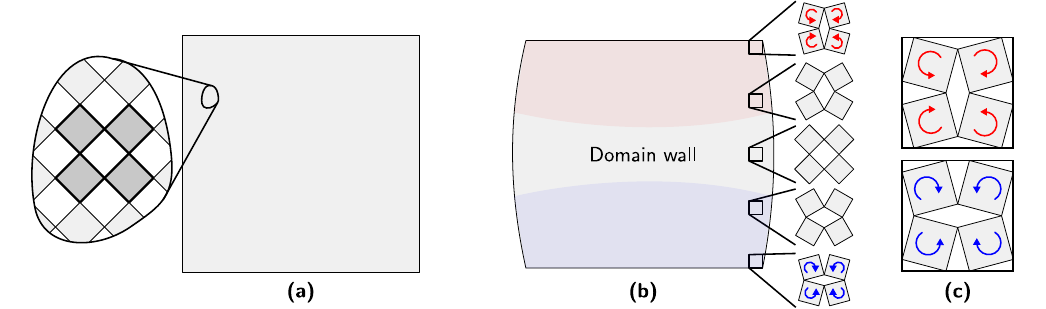}
\caption{Example of ``micro"-incompatibilities, inspired by Ref.~\cite{deng2020characterization}. (a) The reference configuration is the open state of a rotating-squares pattern. (b) Forcing the pattern into opposite ``red" and "blue" phases at the top and bottom leads to a domain wall. (c) Incompatible microstructures can have the same cell-averaged response.}
\label{fig:intro2}
\end{figure}
To illustrate this point, consider the ``domain wall'' drawn in Fig.\;\ref{fig:intro2}, which we base on an experiment in Ref.~\cite{deng2020characterization}. An initially open rotating-squares specimen (Fig.\;\ref{fig:intro2}(a)) has two symmetry-related slit actuations that result in the same cell-averaged deformation, the ``red" and ``blue" phases in Fig.\;\ref{fig:intro2}(b-c). Forcing opposite phases to coexist leads to elastic frustration and to the appearance of a domain wall motif familiar from other materials applications \cite{brown1966magnetoelastic,toledano1987landau,bhattacharya2003microstructure}.  This frustration is due to the microscale incompatibility of the two phases, which are nevertheless macroscopically compatible from the viewpoint of their effective deformations (both phases achieve the same overall shape change in Fig.\;\ref{fig:intro2}(c)). Thus, domain walls cannot be captured with a model based only on cell-averaged deformations. The fix is to introduce an angle into the model that distinguishes between the two slit-actuations of the incompatible phases. In this way, Deng \textit{et al.}\;predict a one-dimensional profile for the slit actuation across the wall, for sufficiently small displacements allowing the two-dimensional character of the wall to be neglected. We embrace the key insight of including the actuation, and incorporate it here into a nonlinear model. %This formula misses the two-dimensional domain wall structure which emerges for large loads.  % Nevertheless, so only qualitative under typical loading conditions, but the added angle field is a key insight. Its inclusion is necessary to account for all the sources of elasticity in kirigami at the continuum scale.

Elastic continuum models with additional fields are known as \textit{generalized elastic continua}. Such models were proposed by the Cosserat brothers in the late 1800s, and later codified in great detail in the works of Eringen (in \cite{eringen2012microcontinuum} and references therein).  Popular incarnations include micropolar and micromorophic elasticity; in general, a microcontinuum model uses auxilliary continuum fields to capture mechanical rearrangements at the microscale with consequences for elasticity at larger scales. For this reason, perhaps, recent  research on metamaterials has made connections to this classical subject \cite{saremi2020topological,nassar2020microtwist,lakes2022extremal}. Nevertheless, the most familiar microcontinuum models do not appear to apply broadly to mechanical metamaterials, whose  micro-motions are generally nonlinear. On a related note, there has been a systematic effort to coarse-grain the linear response of discrete truss structures including the Pantographic lattice \cite{alibert2003truss,abdoul2018strain,seppecher2019pantographic,durand2022predictive}, with the aim of setting strain-gradient and higher order effective continuum models on rigorous grounds. However, these results are limited to small displacements and linear elastic responses.  

A key task in the continuum modeling of mechanism-based metamaterials is to link the (fundamentally nonlinear) micro-motions of the panels in a soft mode to the effective, macro-scale shape change. Our recent work \cite{zheng2022continuum} accomplishes this for planar kirigami with a nonlinear partial differential equation (PDE) relating the micro- and macro-scale motions. However, enforcing this PDE as a purely geometric constraint on the effective deformation neglects higher-order elastic effects present at the micro-scale, which contain information needed to formulate and solve elastic boundary value problems.  We address this issue in this paper by positing a second-gradient like (actually mechanism gradient) generalized continuum model that enables us to solve for the response of the kirigami to general boundary conditions and loads.

Specifically, we present a generalized elastic continuum model for planar kirigami that accounts for the geometric nonlinearity of its soft modes. For concreteness and simplicity, we focus on a model system termed \textit{rhombi-slit} krigami as it features periodic patterns of quad panels and rhombi slits. All such patterns possess a periodic mechanism fully parameterized by a quantity we call the \textit{slit actuation}; all such patterns also exhibit locally mechanistic soft modes, whose effective description is captured by the PDE derived in \cite{zheng2022continuum}. Here, we build the PDE into a constitutive model with a bulk elastic energy that vanishes on its solutions, yielding an effective stress that vanishes on soft modes. The model also includes two higher order sources of elasticity of physical origin \cite{coulais2018characteristic, czajkowski2022conformal}, which we model using the slit actuation field: its gradient accounts for deviations from a pure mechanism, while its value accounts for hinge bending. Altogether, these three terms furnish a generalized elastic continuum model for rhombi-slit krigami in the plane. This new continuum constitutive model provides a versatile framework for solving elastic boundary value problems using a standard FEM platform (Abaqus). We demonstrate this versatility by comparing simulations to experiments across designs and loading conditions. 

 The rest of this paper is organized as follows.  Section \ref{sec:s2} describes the geometry of rhombi-slit kirigami and parameterizes their mechanisms. Section \ref{sec:Model} introduces our generalized elastic continuum model. Section \ref{sec:EquilFEA} derives its equilibrium equations and provides a finite element formulation. Section \ref{sec:s4} compares simulations of the model to experiments; {\color{black}the model reproduces heterogeneous displacement fields as well as a force--displacement curve.} Section \ref{sec:s5} ends with concluding remarks.

\section{Kirigami patterns and their mechanisms}\label{sec:s2}

To introduce the design and kinematic variables of our model, we first treat the kirigami's simplest modes of deformation: its pure mechanisms. 

\subsection{Rhombi-slit kirigami designs}\label{ssec:s21}

\begin{figure}[htb!]
\centering
\includegraphics[scale=1.0]{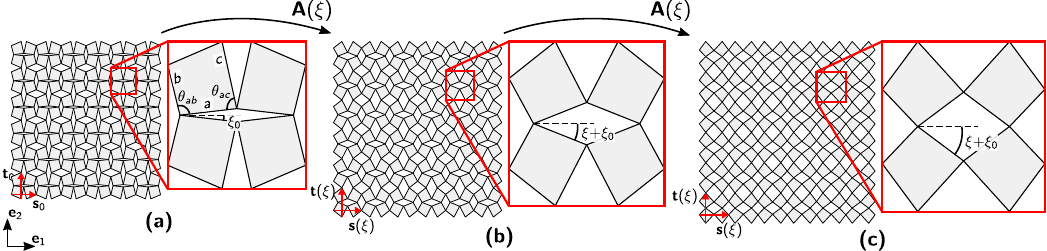}
\caption{Effective deformation of a planar kirigami metamaterial made of periodic arrays of a unit cell featuring four quad panels and four rhombi-slits. (a) Reference configuration, with detail of a unit cell and its characteristic dimensions. Panels have side lengths $a$, $b$, $c$ and internal angles $\theta_{ab}$ and $\theta_{ac}$. The central slit has an initial opening angle $2\xi_0$. (b,c) Instances of the pure mechanism motion of the metamaterial in (a). During the mechanism motion, the unit cell represented by Bravais lattice vectors $\mathbf{s_0}$ and $\mathbf{t_0}$ in (a) is stretched and rotated, according to the stretch tensor $\mathbf{A}(\xi)$, where $2\xi$ is the change of opening angle of the central slit. Correspondingly, the lattice vectors deform to $\mathbf{s}(\xi)$ and $\mathbf{t}(\xi)$.}
\label{fig:eff}
\end{figure}

We consider planar kirigami metamaterials consisting of a periodic array of unit cells, each having four quadrilateral panels and four rhombi-slits, as in Fig.\;\ref{fig:eff}(a). In such designs, the four panels of the cell have identical shape and are mirrors of each other across the slit-axes.  We use edge lengths $a$, $b$, and $c$ and two sector angles $\theta_{ab}$ and $\theta_{ac}$ to describe the shape of the quadrilateral panels. We also denote by $\xi_0$ the half opening angle of the central slit of the cell in its reference configuration. These six parameters are sketched in Fig.\;\ref{fig:eff}(a) and fully parameterize the unit cell, up to a rigid motion.  This family of metamaterials has a large design space: it includes the well-known rotating-squares patterns ($a=b=c$, $\theta_{ab}=\theta_{ac}=\pi/2$) in its closed state $(\xi_0 = 0)$ or open state ($\xi_0 = \pi/4)$.
% 
 
%For one unit cell showed in Fig.\;\ref{fig:eff}, one center slit with side length $a$ is surrounded with four quadrilateral panels. We define the angle change of the center slit as $2\xi$.

\subsection{Effective description of kirigami mechanisms}\label{ssec:EffDefMech}
Each rhombi-slit kirigami possesses a single degree-of-freedom planar mechanism.  We now recall the relevant results from \cite{zheng2022continuum}, which link the kirigami's effective deformation to its slit actuation.

Let $\mathbf{e}_1$ and $\mathbf{e}_2$ denote the standard 2D Cartesian basis. The vectors $\mathbf{s}_0$ and $\mathbf{t}_0$ in Fig.\;\ref{fig:eff}(a) are Bravais lattice vectors reflecting the periodicity of the pattern in its reference configuration. They can be written explicitly in terms of the cell parameters as
\begin{equation}
    \begin{aligned}
    &\mathbf{s}_0 = 2(a \cos{\xi_0} + b \cos{(\pi-\theta_{ab}-\xi_0)} )\mathbf{e}_1, \\
    &\mathbf{t}_0=2( a \sin{\xi_0} + c \cos{(\pi/2-\theta_{ac}+\xi_0)}) \mathbf{e}_2. 
    \end{aligned}
\end{equation}
Now replace $\xi_0$ by $\xi_0 + \xi$ and observe from Fig.\;\ref{fig:eff}(b,c) that this change results in a deformed configuration of the pattern, where the central slit of each cell opens by an angle $2 \xi$ and the panels counter-rotate to accommodate this actuation. In other words, \textit{slit actuation by} $\xi$ parameterizes a mechanism deformation of the pattern. As shown, the deformed configuration has the Bravais lattice vectors 
\begin{equation}
    \begin{aligned}
    &\mathbf{s}(\xi) =2( a \cos{(\xi_0 + \xi)} + b \cos{(\pi-\theta_{ab}-\xi_0 - \xi)}) \mathbf{e}_1, \\
    &\mathbf{t}(\xi)=2( a \sin{(\xi_0 + \xi)} + c \cos{(\pi/2-\theta_{ac}+\xi_0 + \xi)}) \mathbf{e}_2.
    \end{aligned}
\end{equation}

To track the shape-change associated to this mechanism, we introduced a \textit{shape tensor}  $\mathbf{A}(\xi)$. This $2 \times 2$ tensor is defined as the unique linear transformation taking the reference Bravais lattice vectors to their deformed counterparts. It is determined as a function of the slit actuation $\xi$ via
\begin{equation}
    \begin{aligned}
    \mathbf{s}(\xi) = \mathbf{A}(\xi) \mathbf{s}_0, \quad \mathbf{t}(\xi) = \mathbf{A}(\xi) \mathbf{t}_0.
    \end{aligned}
\end{equation}
The explicit formula for $\mathbf{A}(\xi)$ in the case of rhombi-slit kirigami is
\begin{equation}
\begin{aligned}\label{eq:Axi}
&\mathbf{A}(\xi) = (\cos{\xi} -\alpha \sin{\xi})\mathbf{e}_1 \otimes \mathbf{e}_1 + (\cos{\xi} +\beta \sin{\xi})\mathbf{e}_2 \otimes \mathbf{e}_2, 
\\
&\alpha = \frac{a \sin{\xi_0}  - b \sin{( \theta_{ab}+ \xi_0)}}{ a \cos{\xi_0}-b \cos{( \theta_{ab}+\xi_0)} }  ,  \quad \beta = \frac{ -c \cos{( \theta_{ac}- \xi_0)} +a \cos{\xi_0} }{ c \sin{( \theta_{ac}-\xi_0)} + a \sin{\xi_0} }.  
\end{aligned}
\end{equation}
Likewise, to coarse-grain the panel motions we introduced a 2D \textit{effective deformation}  $\mathbf{y}_{\text{eff}}(\mathbf{x})$ %denote a 2D effective deformation, linked to a planar mechanism deformation of the pattern
per the Cauchy-Born rule. 
Since all cells deform identically in a mechanism, the associated $\mathbf{y}_{\text{eff}}(\mathbf{x})$ is homogeneous. Its $2\times 2$ deformation gradient    $\mathbf{F}_{\text{eff}}$ is constant, and is parameterized by the shape tensor up to a rigid rotation. We showed that 
\begin{equation}
\begin{aligned}\label{eq:PureMech}
(\text{pure mechanism:})& \quad (\mathbf{F}_{\text{eff}})^T \mathbf{F}_{\text{eff}} = \mathbf{A}^2(\xi)
\end{aligned}
\end{equation}
thereby quantifying the link between the effective deformation of a mechanism and its slit actuation, $\xi$. By varying $\xi$, one obtains a compact description of the effective shape change.  

Finding the shape tensor of a periodic mechanical metamaterial is a general way of linking its micro-scale motion to its macro-scale deformations, which we have just demonstrated for the pure mechanisms of rhombi-slit kirigami. %This approach can be used to quantify the effective shape change due to a mechanism in \textit{any} periodic mechanical metamaterial with periodic mechanism deformations. 
The same approach applies to the more general family of parallelogram-slit kirigami, and goes beyond its mechanisms to capture its soft modes \cite{zheng2022continuum}. Certain generalized Miura-Ori origami patterns have been similarly coarse-grained in \cite{mcinerney2021discrete,nassar2022strain}. 
We focus on the class of rhombi-slit kirigami here because we think it strikes a balance between breadth and simplicity. Its simplicity is reflected in the fact that its shape tensor is diagonal (Eq.\;(\ref{eq:Axi})), reflecting a locally bi-axial shape change. Its breadth will become clearer as we go on to describe a model for predicting the wealth of non-homogeneous soft modes.

%To obtain the deformation of kirigami metamaterials, we started with the mechanism of the inside kirigami pattern. When the slit of the unit cell opens, the kirigami pattern expands in lateral directions accordingly. Let $\xi_0$ denotes the slit angle in the initial state, the current opening angle $\xi_t=\xi_0+\xi$. Setting basis vectors $\mathbf{e}_1$ and $\mathbf{e}_2$ as the standard 2D Cartesian coordinate, we coarse grain the kirigami pattern with vectors $\mathbf{s_0} = (a \cos{\xi_0} + b \cos{(\pi-\theta_{ab}-\xi_0)} )\mathbf{e}_1$ and $\mathbf{t_0}=( a \sin{\xi_0} + c \cos{(\pi/2-\theta_{ac}+\xi_0)}) \mathbf{e}_2$ in the initial state. Due to the inner slit opens, pattern deforms to $\mathbf{s} =( a \cos{\xi_t} + b \cos{(\pi-\theta_{ab}-\xi_t)}) \mathbf{e}_1$ and $\mathbf{t}=( a \sin{\xi_t} + c \cos{(\pi/2-\theta_{ac}+\xi_t)}) \mathbf{e}_2$. If there exists an tensor $\mathbf{A}$ satisfying that  $\mathbf{s} = \mathbf{A}\mathbf{s_0}$, $\mathbf{t} = \mathbf{A}\mathbf{t_0}$ (Fig.\;\ref{fig:eff}), $\mathbf{A}$ can be regarded as the effective stretch tensor.

\section{A generalized continuum model for planar kirigami}\label{sec:Model}

We now present a constitutive model for the elasticity of rhombi-slit kirigami, based on the effective description of its mechanisms recalled in the previous section. As noted in the introduction, the basic experimental observation is that for a large class of loading conditions, the kirigami exhibits a soft response that cannot be captured by any single mechanism. Instead, its cells deform by an approximately locally mechanistic response, with an actuation that varies slowly from cell to cell, and with panel deformations that appear to oscillate about some smooth continuum deformation. We therefore build a generalized continuum model to predict the deformation and its underlying actuation.

\subsection{Statement of the model}\label{ssec:Model1}

%Our model begins by introducing a bulk elastic term to drive the pattern's effective deformation towards a configuration consistent with cell-averaged local mechanisms. 

For a rhombi-slit kirigami filling a 2D reference domain $\Omega$, let  $\mathbf{y}_{\text{eff}}\colon \Omega \rightarrow \mathbb{R}^2$ denote its effective deformation and $\xi \colon \Omega \rightarrow \mathbb{R}$ its slit actuation, both of which we understand as continuum fields.
Building off of Eq.\;(\ref{eq:PureMech}), our general idea is to choose an energy whose leading order behavior prefers approximate local mechanisms. That is, we will require the model to produce the response
\begin{equation}
    \begin{aligned}\label{eq:softMode}
    (\text{approximately locally mechanistic:}) \quad  \big(\nabla \mathbf{y}_{\text{eff}}(\mathbf{x})\big)^T \nabla \mathbf{y}_{\text{eff}}(\mathbf{x}) \approx \mathbf{A}^2(\xi(\mathbf{x})).
    \end{aligned}
\end{equation}
%with $\mathbf{y}_{\text{eff}}(\mathbf{x})$ and $\xi(\mathbf{x})$ being the effective deformation and slit actuation field, respectively. 
The pattern's slit actuation $\xi$ will be treated as an auxiliary field variable governed by higher-order sources of elasticity in addition to this leading order constraint.

Proceeding to details, we %consider the total potential energy of the system to be a function of these fields.
assume that the fields $(\mathbf{y}_{\text{eff}},\xi)$ describing the kirigami's response to loads are (local) minimizers of the potential energy function
\begin{equation}
    \begin{aligned}\label{eq:PotentialEnergy}
    &\mathcal{E}(\mathbf{y}_{\text{eff}},\xi) = \int_{\Omega} W(\nabla \mathbf{y}_{\text{eff}}(\mathbf{x}), \xi(\mathbf{x}), \nabla \xi(\mathbf{x})) dA - \int_{\partial_t \Omega} \mathbf{t}_{\text{R}}(\mathbf{x}) \cdot \mathbf{y}(\mathbf{x}) d \Gamma,  \\
    &\text{subject to } \mathbf{y}_{\text{eff}}(\mathbf{x}) = \mathbf{y}_{\text{b}}(\mathbf{x}) \text{ on } \partial \Omega \setminus \partial_t \Omega,
    \end{aligned}
\end{equation}
where $\mathbf{t}_{\text{R}}$ denotes a prescribed ``reference" traction on the boundary component $\partial_t \Omega \subset \partial \Omega$, and $\mathbf{y}_{\text{b}}$ prescribes the deformation on the rest of the boundary. We use the strain energy density  
\begin{equation}
    \begin{aligned}\label{eq:strainEnergy}
    W(\mathbf{F}, \theta, \mathbf{p}) = c_0 W_0( \mathbf{F} \mathbf{A}^{-1}(\theta)) + c_1  \theta^2 + c_2 |\mathbf{p}|^2.
    \end{aligned}
\end{equation}
Its first term, $W_0$,  takes the form of a standard isotropic 2D hyperelastic model 
\begin{equation}
    \begin{aligned}\label{eq:hookean}
    W_0(\mathbf{G}) =\Big(\frac{1}{J_{\mathbf{G}}} |\mathbf{G}|^2 -2\Big) + (J_{\mathbf{G}} - 1)^2
    \end{aligned}
\end{equation}
up to the decomposition $\mathbf{G} = \mathbf{F} \mathbf{A}^{-1}(\theta)$. Here $\mathbf{A}(\theta)$ is the shape tensor, encoding the geometry of the reference pattern from Section \ref{ssec:s21}; it is invertible for any physical value of slit actuation. Note $J_{\mathbf{G}} = \det \mathbf{G}$ for short. Finally, the parameters $c_0,c_1,c_2$ are elastic moduli which we anticipate fitting to experiments. 

\subsection{Physical origin of the model}\label{ssec:Model-physical-origin}

\begin{figure}[htb!]
\centering
\includegraphics[scale=1.0]{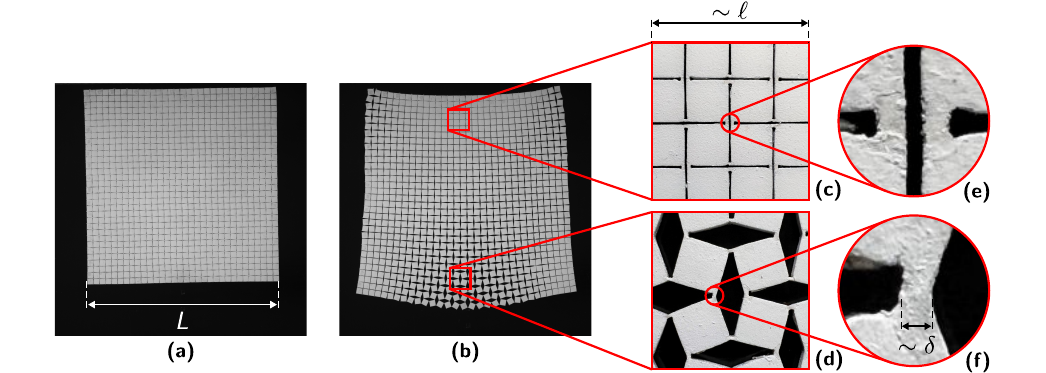}
\caption{Illustration of the characteristic lengthscales of a rhombi-slit kirigami metamaterial. (a) Reference, undeformed configuration, indicating the sample lengthscale $L$. (b) Same pattern, subjected to heterogeneous loading. (c,d) Zoom-in on slightly deformed and significantly deformed regions of the specimen, through a window of dimensions comparable to the unit cell lengthscale $\ell$. (e,f) Zoom-in on the hinges, indicating the hinge lengthscale $\delta$.}
\label{fig:MacroToMicro}
\end{figure}

Using Fig.\;\ref{fig:MacroToMicro} as a guide, we now discuss the physical origin of the three terms in the energy density.  Each term is a distinct and natural consequence of the disparity of lengthscales in the pattern. As sketched in Fig.\;\ref{fig:MacroToMicro}(a), (c,d) and (e,f), $L$ is the sample lengthscale, $\ell$ the unit cell lengthscale, and $\delta $ the hinge lengthscale. Though our model can be generally applied, it is most relevant when $\delta \ll \ell \ll L$, i.e., when the number of unit cells in the sample is large, and the hinges are small as compared to the panels. We make this assumption throughout.

The first term in Eq.\;(\ref{eq:strainEnergy}), $W_0$, arises as follows. Suppose we focus solely on regions of length $\sim \ell$ under typical loading, like those in Fig.\;\ref{fig:MacroToMicro}(c,d). The local response is then nearly indistinguishable from that of a pure mechanism; the heterogeneity, as seen in Fig.\;\ref{fig:MacroToMicro}(b), only emerges at a much larger scale. Because of this feature, we choose $W_0$ to be a bulk elastic energy that is minimized and zero whenever $(\mathbf{y}_{\text{eff}},\xi)$ coincide with a coarse-grained local mechanism,
\begin{equation}
    \begin{aligned}\label{eq:softMode-exact}
    (\text{locally mechanistic:}) \quad  \big(\nabla \mathbf{y}_{\text{eff}}(\mathbf{x})\big)^T \nabla \mathbf{y}_{\text{eff}}(\mathbf{x}) = \mathbf{A}^2(\xi(\mathbf{x})).
    \end{aligned}
\end{equation}
Choosing $W_0$ as such introduces stress proportional to the modulus $c_0$ to penalize deviations from local mechanisms. This stress should be significant. In fact, as was described in the supplementary material of our previous work \cite{zheng2022continuum}, a coarse-grained kirigami deformation that fails to be locally mechanistic is accompanied by distortions of the panels and hinges comparable to the characteristic length of the panels ($\sim \ell)$. As such, $c_0$ should scale with the shear modulus $\mu$ of the material used to fabricate the kirigami (possibly with some weak dependence on $\ell/L$ or $\delta/\ell$).

The other two terms in the energy density Eq.\;(\ref{eq:strainEnergy}) account for higher order effects related to localized elastic distortions in and nearby the hinges. The second term (with modulus $c_1$) is the simplest possible model that approximates the bending energy of the hinges when slits open or close. In the effective description, it  depends only on the slit actuation angle $\xi$. This actuation counter-rotates the panels to yield the bending, as shown in Fig.\;\ref{fig:MacroToMicro}(d,f). Large actuation causes localized strains {\color{black}$\sim |\xi|$} nearby the hinges, {\color{black} which give a bending energy per cell $\sim \mu |\xi|^2\delta^2$ since the hinges occupy an area $\sim \delta^2$. As the cell area is $\sim \ell^2$, the corresponding modulus $c_1$ in Eq.\;(\ref{eq:strainEnergy}) scales as $c_1 \sim \mu \delta^2/\ell^2$}.

The final term in the energy density Eq.\;(\ref{eq:strainEnergy}) regularizes the actuation field, and corresponds to the simplest elastic model for resisting inhomogeneous actuation in a locally mechanistic response ({\color{black}i.e., ``mechanism gradients''}). This term is akin to the strain-gradient term in Ref.~\cite{czajkowski2022conformal} and the plate bending term from Ref.~\cite{MCMAHAN2022}, {\color{black}which respectively regularize mechanism gradients via second gradients of the effective in- and out-of-plane displacements. In contrast, our model depends directly on the actuation gradient $\nabla \xi$. As such, it accounts for all possible micro-incompatibilities, including those that lead to a homogeneous effective displacement, and that would be spuriously assigned no strain-gradient/plate bending energy. \color{black}Fig.\;\ref{fig:intro2} from the introduction gives an example.} 

Our mechanism gradient term also has physical origins \cite{coulais2018characteristic}. Where the kirigami's response deviates from a single mechanism, elastic distortions are required to preserve the connectivity of the pattern. {\color{black}For a locally mechanistic deformation with a smoothly varying actuation field $\xi$, one expects the required distortions to be $\sim \ell^2|\nabla \xi|$ in magnitude, and to be accompanied by panel strains $\sim \ell|\nabla \xi|$ \cite{zheng2022continuum}. (In fact, there is a subtle logarithmic correction to these scalings involving a self-similar spreading of strains which will be the topic of forthcoming work \cite{plucinskyetal}.)  This argument gives an energy per cell $\sim \mu \ell^2 |\nabla \xi|^2\ell^2$ since each cell has area $\sim \ell^2$. The corresponding modulus $c_2$ in Eq.\;(\ref{eq:strainEnergy}) scales as $c_2 \sim \mu \ell^2$. Postponing the detailed derivation of the fully tensorial mechanism gradient term to future work \cite{plucinskyetal},} we simply choose to model it here using a straightforward constitutive model that contains all the necessary sources of elasticity.

\subsection{Effective stress and effectively stress-free configurations}\label{ssec:StressSec}
Having written down an effective model, we proceed to determine the effective stress measures it imposes, and derive and discuss some properties of effectively stress-free configurations. The first Piola-Kirichhoff stress $\mathbf{P}$ is obtained by differentiating the strain energy density $W$ in $\mathbf{F}$. From Eq.\;(\ref{eq:strainEnergy}), 
\begin{equation}
\begin{aligned}
\mathbf{P}(\mathbf{F},\theta) = \mathbf{P}_0\big(\mathbf{F} \mathbf{A}^{-1}(\theta)\big) \mathbf{A}^{-T}(\theta)
\end{aligned}
\end{equation}
for the 2D hyperelastic first Piola-Kirchhoff stress
\begin{equation}
    \begin{aligned}
    \mathbf{P}_0(\mathbf{G}) := c_0\frac{\partial W_0(\mathbf{G})}{\partial \mathbf{G}} =  \frac{ c_0}{J_{\mathbf{G}}} \Big(  2\mathbf{G} + \big(2(J_{\mathbf{G}}^3 -J_{\mathbf{G}}^2)  - |\mathbf{G}|^2 \big) \mathbf{G}^{-T} \Big) .
    \end{aligned}
\end{equation}
We also introduce the so-called Kirchhoff stress \cite{bower2009applied}, as it will be useful later on for numerics. This stress is defined as
\begin{equation}
    \begin{aligned}
    \boldsymbol{\tau}(\mathbf{F}, \theta) := \mathbf{P}(\mathbf{F},\theta) \mathbf{F}^T  =  \mathbf{P}_0(\mathbf{F} \mathbf{A}^{-1}(\theta)) \big(\mathbf{F} \mathbf{A}^{-1}(\theta)\big)^T =  \boldsymbol{\tau}_0\big(\mathbf{F} \mathbf{A}^{-1}(\theta)\big)
    \end{aligned}
\end{equation}
for the 2D hyperelastic Kirchhoff stress 
\begin{equation}
    \begin{aligned}\label{eq:kirchhoffTensor}
 \boldsymbol{\tau}_0(\mathbf{G}) := \mathbf{P}_0(\mathbf{G}) \mathbf{G}^T =   \frac{ c_0}{J_{\mathbf{G}}} \Big(  2\mathbf{G}\mathbf{G}^T + \big(2(J_{\mathbf{G}}^3 -J_{\mathbf{G}}^2)  - |\mathbf{G}|^2 \big)\mathbf{I} \Big).  
    \end{aligned}
\end{equation}

In a traditional elastic continuum, the stress-free configurations are rigid body motions. Indeed,  $\boldsymbol{\tau}_0(\mathbf{G}) = \mathbf{0}$ if and only if $\mathbf{G}$ is a rotation. In contrast, our generalized continuum model has a much richer family of effectively stress-free configurations: for the fields $(\mathbf{y}_{\text{eff}}, \xi)$ on $\Omega$,
\begin{equation}
    \begin{aligned}\label{eq:StressFree1}
    \boldsymbol{\tau}(\nabla \mathbf{y}_{\text{eff}}(\mathbf{x}), \xi(\mathbf{x})) = \mathbf{0} \quad \Leftrightarrow \quad \big(\nabla \mathbf{y}_{\text{eff}}(\mathbf{x})\big)^T \nabla \mathbf{y}_{\text{eff}}(\mathbf{x})  =  \mathbf{A}^2(\xi(\mathbf{x})) .
    \end{aligned}
\end{equation} 
The latter equation is the condition for a coarse-grained local mechanism discussed above (Eq.\;(\ref{eq:softMode-exact})).  It constrains the  metric tensor (right Cauchy-Green tensor) of the effective deformation $\mathbf{y}_{\text{eff}}$ in terms of the slit actuation field $\xi$, which reflects the fact that the effective fields must vary in a coordinated way to represent a locally mechanistic soft mode of rhombi-slit kirigami in the plane.

In our previous work \cite{zheng2022continuum}, we showed that there are broad classes of deformations and slit actuations that  solve this metric constraint, and that these fields capture the behavior of soft deformations observed experimentally.  We also derived an important link between  the metric constraint and the effective Poisson's ratio of the pattern. We present a brief description of this link here, as it illuminates a fundamental dichotomy in the qualitative features of the slit actuation in this model, and will help to clarify the choices we make later on in the simulations and experiments.

Consider any effectively stress-free configuration $(\mathbf{y}_{\text{eff}}, \xi)$, i.e., one with $\boldsymbol{\tau}=\mathbf{0}$. %that satisfies the metric constraint $\big(\nabla \mathbf{y}_{\text{eff}}(\mathbf{x})\big)^T \nabla \mathbf{y}_{\text{eff}}(\mathbf{x}) = \mathbf{A}^2(\xi(\mathbf{x}))$.
Since the deformation is planar, it trivially has zero Gauss curvature. As the Gauss curvature is fundamentally linked to its  metric tensor through Gauss's remarkable theorem \cite{do2016differential}, we find that the slit actuation solves a PDE of the form
\begin{equation}
    \begin{aligned}\label{eq:slitPDE}
    \Big[\partial_2^2  - \big(\tfrac{\mu_1(\xi(\mathbf{x}))}{\mu_2(\xi(\mathbf{x}))}\big)^2 \nu_{21}(\xi(\mathbf{x})) \partial_1^2\Big]  \xi(\mathbf{x})  = g(\xi(\mathbf{x}), \nabla \xi(\mathbf{x})) ,
    \end{aligned}
\end{equation}
for $g(\theta, \mathbf{p})$ a function whose explicit form is lengthy and not important for the discussion at hand.  The function $\nu_{21}(\theta)$, introduced in this PDE,  turns out to be the effective Possion's ratio of the pattern.  It satisfies 
\begin{equation}
    \begin{aligned}\label{eq:PRatio}
    \nu_{21}(\theta)  = \frac{(\cos \theta - \alpha \sin \theta)}{(\sin \theta + \alpha \cos \theta)} \frac{(-\sin \theta + \beta \cos \theta)}{(\cos \theta + \beta \sin \theta)}
    \end{aligned}
\end{equation}
where  $\alpha$ and $\beta$ quantify the reference rhombi-slit cell via Eq.\;(\ref{eq:Axi}). 

The PDE in Eq.\;(\ref{eq:slitPDE}) is second order and quasi-linear, which means that its so-called \textit{type} has a standard classification  from  PDE theory \cite{courant2008methods}: the PDE is  elliptic for solutions $\xi$ that satisfy $\nu_{21}(\xi(\mathbf{x})) <0$ on $\Omega$, is hyperbolic for solutions that satisfy $\nu_{21}(\xi(\mathbf{x})) > 0$ on $\Omega$, and is of mixed type if the solution's Poisson's ratio changes sign on some part of the domain.  In other words, the PDE is elliptic if the unit cell is auxetic and hyperbolic if it is \textit{non}-auxetic. 

These results have important modeling implications. In the typical setting, where $c_0 \gg c_{1,2}$ in Eq.\;(\ref{eq:strainEnergy}), we expect a large family of  ``slightly stressed" equilibrium solutions of our model. Each such solution, then, approximately solves the metric constraint, with a $\xi$ field that closely resembles one obeying the PDE in Eq.\;(\ref{eq:slitPDE}).  Given this observation, we identify rhombi-slit kirigami as elliptic or hyperbolic based on the sign of its effective Poisson's ratio at equilibrium.  We expect this identification to indicate certain qualitative properties of the kirigami's response to loads:  elliptic kirigami with auxetic cells should exhibit a decay in slit actuation away from the boundary loads. In contrast, hyperbolic kirigami with non-auxetic cells should exhibit comparatively persistent slit actuation under the same loading conditions. We highlight this   dichotomy in further detail with the examples to come.

\section{Equilibrium equations and finite element formulation}\label{sec:EquilFEA}

Next, we derive the equilibrium equations and natural boundary conditions of our model.  The equilibrium equations are a coupled PDE system that includes the familiar divergence-free condition on the effective stress, as well as an auxiliary PDE enforcing equilibrium for the slit actuation. We finish this section with an FEM formulation of the model, and discuss its implementation in  Abaqus.

\subsection{Derivation of equilibrium equations}
We first derive the equilibrium equations. Let $(\mathbf{y}_{\text{eff}}, \xi)$ be a local minimizer to the potential energy in Eq.\;(\ref{eq:PotentialEnergy}). Then, by taking the first variation of the energy functional,
\begin{equation}
    \begin{aligned}\label{eq:firstVariation}
    0 &= \frac{d}{d \epsilon} \mathcal{E}(\mathbf{y}_{\text{eff}} + \epsilon \mathbf{w}, \xi +  \epsilon  \eta) \big|_{\epsilon = 0} \\
    &= \int_{\Omega}  \mathbf{P}\big(\nabla \mathbf{y}_{\text{eff}}(\mathbf{x}), \xi(\mathbf{x}) \big) \colon \nabla \mathbf{w}(\mathbf{x})  dA - \int_{\partial_t \Omega} \mathbf{t}_{\text{R}}(\mathbf{x}) \cdot \mathbf{w}(\mathbf{x})  d\Gamma  \\
    &\quad + \int_{\Omega}\Big\{ W_{\theta}\big(\nabla \mathbf{y}_{\text{eff}}(\mathbf{x}), \xi(\mathbf{x}), \nabla \xi(\mathbf{x})\big) \eta(\mathbf{x}) +  W_{\mathbf{p}}\big(\nabla \mathbf{y}_{\text{eff}}(\mathbf{x}), \xi(\mathbf{x}), \nabla \xi(\mathbf{x})\big) \cdot \nabla \eta(\mathbf{x}) \Big\}  dA 
    \end{aligned}
\end{equation}
for all sufficiently smooth $\mathbf{w} \colon \Omega \rightarrow \mathbb{R}^2$ satisfying $\mathbf{w}(\mathbf{x}) = \mathbf{0}$ on $\partial \Omega \setminus \partial_t \Omega$, and $\eta \colon \Omega \rightarrow \mathbb{R}.$ Here and throughout, we use a colon to denote the contraction of square matrices $\mathbf{S} \colon \mathbf{T} = \Tr(\mathbf{S}^T \mathbf{T})$. In addition,  $W_{\theta}$ and $W_{\mathbf{p}}$ denote the partial derivatives of $W(\mathbf{F}, \theta, \mathbf{p})$ with respect to $\theta$ and $\mathbf{p}$. Observe that 
\begin{equation}
    \begin{aligned}\label{eq:goodFormula}
    &W_{\theta}(\mathbf{F}, \theta, \mathbf{p}) = \underbrace{-\mathbf{P}_0(\mathbf{F} \mathbf{A}^{-1}(\theta)) \colon \big(\mathbf{F} \mathbf{A}^{-1}(\theta) \mathbf{A}'(\theta) \mathbf{A}^{-1}(\theta)  \big)  + 2c_1 \theta}_{=: f_{\text{act}}(\mathbf{F}, \theta)} , \\
    &W_{\mathbf{p}}(\mathbf{F}, \theta, \mathbf{p}) = 2c_2 \mathbf{p}.
    \end{aligned}
\end{equation}
We show below that $f_{\text{act}}(\mathbf{F},\theta)$ provides a driving force for heterogeneous actuation within the model. 

We substitute these formulas into Eq.\;(\ref{eq:firstVariation}), let $\mathbf{P}_{\mathbf{y}_{\text{eff}},\xi}(\mathbf{x}):= \mathbf{P}(\nabla \mathbf{y}_{\text{eff}}(\mathbf{x}), \xi(\mathbf{x}))$ and apply the divergence theorem with the usual localization arguments to conclude the following equilibrium equations on $\Omega$: 
\begin{equation}
    \begin{aligned}
    \begin{cases}\label{eq:strongForm}
    \nabla \cdot \big(\mathbf{P}^T_{\mathbf{y}_{\text{eff}},\xi}(\mathbf{x}) \big)  = \mathbf{0} \\
    f_{\text{act}}\big(\nabla \mathbf{y}_{\text{eff}}(\mathbf{x}), \xi(\mathbf{x})\big)  = 2c_2 \nabla^2 \xi(\mathbf{x}).
    \end{cases}
    \end{aligned}
\end{equation}
The terminology for $f_\text{act}$ can now be understood, since the actuation at equilibrium is heterogeneous $(\nabla \xi \neq 0)$ whenever $f_{\text{act}}$ is non-zero.  Likewise, we derive the boundary conditions 
\begin{equation}
    \begin{aligned}
    \begin{cases}\label{eq:strongForm2}
    \mathbf{P}_{\mathbf{y}_{\text{eff}},\xi}(\mathbf{x}) \mathbf{n}_{\text{R}}(\mathbf{x}) = \mathbf{t}_{\text{R}}(\mathbf{x}) & \text{ on $\partial_t \Omega$} \\
    \mathbf{y}_{\text{eff}}(\mathbf{x}) = \mathbf{y}_{\text{b}}(\mathbf{x}) & \text{ on  $\partial \Omega \setminus \partial_t \Omega$} \\
    \nabla \xi(\mathbf{x}) \cdot \mathbf{n}_{\text{R}}(\mathbf{x}) =  0  & \text{ on $\partial \Omega$}
    \end{cases}
    \end{aligned}
\end{equation}
where $\mathbf{n}_{\text{R}}(\mathbf{x})$ denotes the outwards-pointing unit normal at a point $\mathbf{x}$ on the boundary of $\Omega$.

\subsection{Finite element formulation}\label{ssec:FEM}

To solve these equations, we use the finite element method (FEM). Eqs.\;(\ref{eq:strongForm}-\ref{eq:strongForm2}) give the strong form of the boundary value problem for the effective deformation $\mathbf{y}_{\text{eff}}$ and slit actuation field $\xi$. FEM formulations are instead based on the weak form, in Eq.\;(\ref{eq:firstVariation}). Using  Eq.\;(\ref{eq:goodFormula}), we can conveniently rewrite the weak form as
\begin{equation}
\begin{aligned}\label{eq:WeakForm}
&\int_{\Omega} \mathbf{P}(\nabla \mathbf{y}_{\text{eff}}(\mathbf{x}), \xi(\mathbf{x})) : \nabla \mathbf{w}(\mathbf{x}) dA- \int_{\partial_t \Omega} \mathbf{t}_{\text{R}}(\mathbf{x}) \cdot \mathbf{w}(\mathbf{x}) d\Gamma =  \mathbf{0},
\\
&\int_{ \Omega } \big\{ f_{\text{act}}\big(\nabla \mathbf{y}_{\text{eff}}(\mathbf{x}), \xi(\mathbf{x})\big)\eta(\mathbf{x}) + 2 c_2 \nabla \xi(\mathbf{x}) \cdot \nabla \eta(\mathbf{x}) \big\}   dA=0
\\
\end{aligned}
\end{equation}
for all sufficiently smooth $\mathbf{w}(\mathbf{x})$ with $\mathbf{w}(\mathbf{x}) = \mathbf{0}$ on $\partial \Omega \setminus \partial_t \Omega$ and all  $\eta(\mathbf{x})$. 

To numerically approximate a solution of Eq.\;(\ref{eq:WeakForm}), we mesh $\Omega$ and set 
\begin{equation}
    \begin{aligned}\label{eq:FEMFields}
    &\mathbf{y}_{\text{eff}}(\mathbf{x}) = \mathbf{x} + \sum_{A} \mathbf{u}^A N^A(\mathbf{x}), && \mathbf{w}(\mathbf{x}) = \sum_{A} \mathbf{w}^A N^{A}(\mathbf{x}),  \\
    &\xi(\mathbf{x}) =  \sum_{A}  \xi^A N^A(\mathbf{x}), && \eta(\mathbf{x}) = \sum_{A} \eta^A N^{A}(\mathbf{x}).
    \end{aligned}
\end{equation}
The quantities $\mathbf{u}^A, \mathbf{w}^A \in \mathbb{R}^2,$ and $\xi^A ,\eta^A \in \mathbb{R}$ denote the nodal values of the respective fields at the mesh points $\mathbf{x}^A \in \Omega$, indexed by $A$; 
$N^A(\mathbf{x})$ is the corresponding shape function with standard properties. 
For organizational purposes, we list all the nodal values $\mathbf{u}^A$ and $\xi^A$ as arrays $\mathbf{u}$ and $\boldsymbol{\xi}$.

We substitute Eq.\;(\ref{eq:FEMFields}) into Eq.\;(\ref{eq:WeakForm}) and use that each $\eta^A$ is arbitrary and each $\mathbf{w}^A$ is arbitrary for $\mathbf{x}^A \notin \partial \Omega \setminus \partial_t \Omega$. After some standard manipulations, we  conclude that $\mathbf{u}$ and $\boldsymbol{\xi}$ should ideally satisfy
\begin{equation}
    \begin{aligned}
    &\mathbf{r}_{\text{force}}^A(\mathbf{u},\boldsymbol{\xi})  = \mathbf{0}  && \text{for all $A$ such that $\mathbf{x}^A \notin \partial \Omega \setminus \partial_t \Omega$},  \\
    &r_{\text{act}}^{A}(\mathbf{u}, \boldsymbol{\xi} )  = 0  && \text{for all $A$}.
    \end{aligned}
\end{equation}
The residual force and actuation are given by 
\begin{equation}
    \begin{aligned}\label{eq:residuals}
    &\mathbf{r}_{\text{force}}^A(\mathbf{u},\boldsymbol{\xi}) = -   \int_{\Omega}  \mathbf{P}\big(\tilde{\mathbf{F}}(\mathbf{x}, \mathbf{u}), \tilde{\theta}(\mathbf{x}, \boldsymbol{ \xi}) \big) \nabla N^{A}(\mathbf{x}) dA   + \int_{\partial_t \Omega} \mathbf{t}_{\text{R}}(\mathbf{x})N^{A}(\mathbf{x}) d\Gamma ,  \\
    &r_{\text{act}}^A(\mathbf{u}, \boldsymbol{\xi} )  = -\int_{\Omega}\Big\{  f_{\text{act}}\big( \tilde{\mathbf{F}}(\mathbf{x}, \mathbf{u}), \tilde{\theta}(\mathbf{x}, \boldsymbol{\xi}) \big)  N^{A}(\mathbf{x})  + 2c_2 \tilde{\mathbf{p}}(\mathbf{x}, \boldsymbol{\xi}) \cdot \nabla N^A(\mathbf{x}) \Big\}   dA
    \end{aligned}
\end{equation}
and the arguments for $\mathbf{P}(\cdot)$, $f_{\text{act}}(\cdot)$, and so on in these formula  are defined as  
\begin{equation}
    \begin{aligned}
    &\tilde{\mathbf{F}}(\mathbf{x},  \mathbf{u}) := \mathbf{I} + \sum_{B} \mathbf{u}^{B} \otimes \nabla N^{B}(\mathbf{x}), \quad \tilde{\theta}(\mathbf{x}, \boldsymbol{\xi}) := \sum_{B}\xi^B N^B(\mathbf{x}), \quad \tilde{\mathbf{p}}(\mathbf{x}, \boldsymbol{\xi}) :=  \sum_B \xi^B \nabla N^{B}(\mathbf{x}).
    \end{aligned}
\end{equation}

In practice, we do not solve for the arrays $(\mathbf{u}, \boldsymbol{\xi})$ by making the residuals vanish. Instead, a Newton-Raphson scheme is employed in the FEM solver, which chooses nodal quantities by iterative linearization in an effort to make the residuals sufficiently small. The solver in Abaqus takes as input explicit functions of the derivatives of the residuals, i.e., the tangents
\begin{equation}
    \begin{aligned}\label{eq:stiffnessMeasures}
    &\mathbf{K}^{AB}_{\text{disp}}(\mathbf{u}, \boldsymbol{\xi}) := -\frac{\partial  \mathbf{r}_{\text{force}}^{A} (\mathbf{u} ,\boldsymbol{\xi})}{\partial  \mathbf{u}^B},  &&  \mathbf{k}^{AB}_{\text{disp}}(\mathbf{u}, \boldsymbol{\xi}) := -\frac{\partial r_{\text{act}}^{A} (\mathbf{u} ,\boldsymbol{\xi})} {\partial  \mathbf{u}^B}, \\
    &k^{AB}_{\text{act}}(\mathbf{u}, \boldsymbol{\xi})  := -\frac{\partial r_{\text{act}}^{A} (\mathbf{u}, \boldsymbol{\xi})} {\partial \xi^B}, &&\mathbf{k}_{\text{act}}^{AB}(\mathbf{u}, \boldsymbol{\xi})  := -\frac{\partial \mathbf{r}_{\text{force}}^{A} (\mathbf{u}, \boldsymbol{\xi})} {\partial \xi^B}.
    \end{aligned}
\end{equation}
Explicit formulas for these tangents are in \ref{sec:Appendix}.

Eq.\;(\ref{eq:residuals}) and the formulas for Eq.\;(\ref{eq:stiffnessMeasures}) are integrals calculated numerically using Gaussian quadrature. The formulas for the shape functions, Gaussian quadrature, and the surface traction term are all standard, and not discussed here. We  implement this FEM formulation into  Abaqus using a user-element subroutine (UEL) \cite{0b112d0e5eba4b7f9768cfe1d818872e}. The UEL is called for each element for each iteration.  The initial nodal coordinates and current nodal variables ($\mathbf{u},\boldsymbol{\xi}$) are input. The nodal residuals in Eq.\;(\ref{eq:residuals}) and tangents in Eq.\;(\ref{eq:stiffnessMeasures}) are output. The UEL can be used for linear/quadratic and triangular/quadrilateral elements. % Our code for the subroutine is available online as supplementary material.

\section{Experiments and simulations}\label{sec:s4}

To validate our model, we compare its predictions with experiments for two designs, namely, the auxetic rotating-squares sample and another that is strictly non-auxetic and is described just below.  We define the geometry of the samples, including hinge regions, and explain how to extract a representative rhombi-slit cell for each design to facilitate a reasonable comparison between model and experiment. We study the simple example of uniform stretch, and then turn to complex examples involving heterogeneous loads.

\subsection{Experimental samples and their idealized unit cells}\label{ssec:ss41}

\begin{figure}[htb!]
\centering
\includegraphics[scale=1.0]{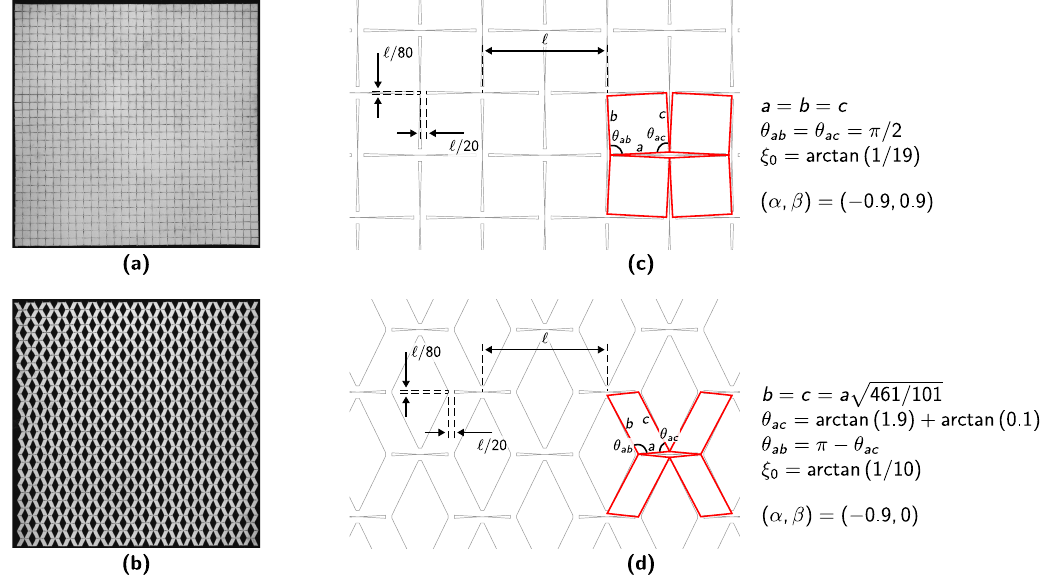}
\caption{From fabricated patterns to idealized rhombi-slit cells. (a) Rotating-squares sample (elliptic). (b) Non-auxetic sample (hyperbolic). (c,d) Portions of the patterns fed to the laser cutter to fabricate specimens (a) and (b). To avoid issues during cutting, hinges are shaped as rectangular blocks. We highlight idealized unit cells in red and report their dimensions.}
\label{fig:LaserCut}
\end{figure}

Fig.\;\ref{fig:LaserCut}(a,b) shows the two specimens used in our work. One is a classical rotating-squares pattern.  The other --- which we term as a ``hyperbolic pattern" for reasons explained below --- features two alternating slits with the same width in the horizontal direction and different height. Fig.\;\ref{fig:LaserCut}(c,d) shows portions of the patterns that were used for laser cutting; the fabricated samples include hinge regions designed to be of length $d = \ell/20$ and  height $h = \ell/80$, as shown.  The actual hinge dimensions vary from the designed ones due to variability associated with laser cutting rubber.

To convert the two cut pattern designs into idealized rhombi-slit designs, we treat the centroid of each hinge as a corner point. Tracing out the lines  connecting neighboring corner points leads to a desired rhombi-slit pattern. We use these ``traced" rhombi-slit cells as inputs to the simulations.  The parameters for both patterns are displayed in Fig.\;\ref{fig:LaserCut}. Recalling Eq.\;(\ref{eq:Axi}), the rotating-squares pattern satisfies $(\alpha, \beta) = (-0.9, 0.9)$, while the hyperbolic pattern satisfies $(\alpha, \beta) = (-0.9, 0)$. Substituting these parameters into Eq.\;(\ref{eq:PRatio}) for the cell's effective Poisson's ratio gives
\begin{equation}
    \begin{aligned}
    &(\text{Rotating-Squares:})  && \nu_{21}(\xi) = -1, \\
    &(\text{Hyperbolic Pattern:}) &&  \nu_{21}(\xi) =  \Big(\frac{-\cos \xi + 0.9 \sin \xi}{-0.9 \cos \xi + \sin \xi}\Big) \tan \xi \overset{|\xi| \ll 1}{\approx} 1.11 \xi.
    \end{aligned}
\end{equation}
As discussed in Section \ref{ssec:StressSec}, the Poisson's ratio describes important qualitative features of each pattern's response, in addition to its auxeticity.

The rotating-squares pattern is auxetic and, more specifically, purely dilational, since the Poisson's ratio indicates equal principal strains independent of the slit actuation. It therefore belongs to the special class of patterns known as \textit{conformal} \cite{czajkowski2022conformal,zheng2022continuum}, which is a subclass of the more general elliptic kirigami discussed previously. We expect this pattern's actuation to generally decay away from boundary loads. In contrast, the hyperbolic pattern is non-auxetic, thus \textit{hyperbolic}, for $\xi \in (0,0.23 \pi)$.  The sample is termed hyperbolic because the slit actuation  is always observed to be in the hyperbolic range for tension type boundary conditions. (In contrast, compressing the sample can lead to buckling, which we do not discuss here; see \cite{MCMAHAN2022} for ideas in this direction.) Guided by its hyperbolicity, we expect this pattern's actuation to persist far away from boundary loads. 

\subsection{Uniform stretch}\label{ssec:ss42}

\begin{figure}[htb!]
\centering
\includegraphics[scale=1.0]{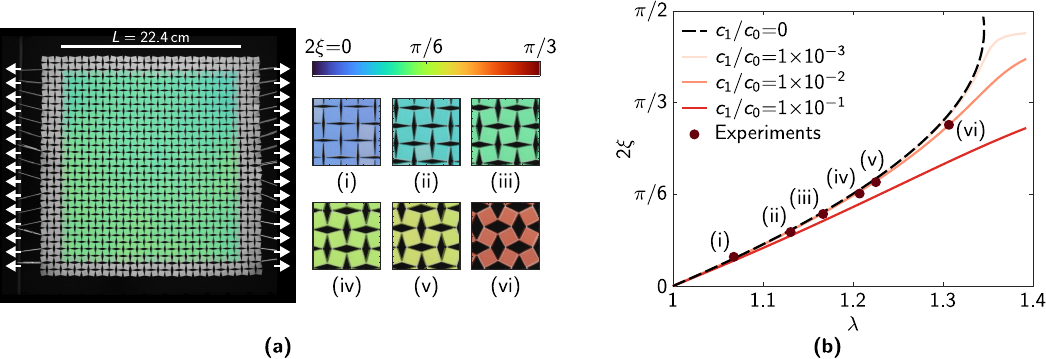}
\caption{Uniform stretch of the rotating-squares sample. (a) Color maps of the slit actuation, determined by the method described in \ref{ssec:Experimental}. $L$ is the undeformed sample length. (b) Slit actuation $\xi$ plotted as a function of horizontal stretch $\lambda$.  Solid lines display the model results with varying $c_1/c_0$.}
\label{fig:fig3}
\end{figure}

The simple loading condition of uniform stretch allows us to investigate the interplay between hinge elasticity and the purely mechanistic response,  both experimentally and in the elastic energy in Eq.\;(\ref{eq:strainEnergy}).  For the demonstration, we focus on the auxetic rotating-squares example.  Experimentally, similarly to what is done in Ref.~\cite{celli2018shape}, we induce free transverse expansion under uniaxial loading by connecting the boundary slits on the left and right edges of the specimen to a rod by means of hooks, as illustrated in  Fig.\;\ref{fig:fig3}(a). This connection allows transverse motion at little resistance. The experimental results, comparing stretch to slit actuation, are marked with circles in {\color{black} Fig.\;\ref{fig:fig3}(b)}. 

The experimental pattern displays a mostly uniform actuation under this loading, especially in the center of the sample as indicated. We therefore consider the elastic energy of the $(\alpha, \beta) = (-0.9,0.9)$ rotating-squares pattern under a homogeneous effective deformation gradient $\nabla \mathbf{y}_{\text{eff}}(\mathbf{x}) = \mathbf{F}(\lambda,\lambda_2) = \lambda \mathbf{e}_1 \otimes \mathbf{e}_1 + \lambda_2 \mathbf{e}_2 \otimes \mathbf{e}_2$ and slit actuation $\xi(\mathbf{x})= \xi$. For a given stretch $\lambda,$ equilibrium is achieved by minimizing $W(\mathbf{F}(\lambda, \lambda_2), \xi, \mathbf{0})$ with respect to  $\lambda_2$ and $\xi$. Setting $w(\lambda, \xi) := \min_{\lambda_2} W(\mathbf{F}(\lambda, \lambda_2), \xi, \mathbf{0})$, the slit actuation at equilibrium  is
\begin{equation}
    \begin{aligned}
    \xi(\lambda) = \arg \min_{\xi}\big\{ w(\lambda, \xi) \big\}.
    \end{aligned}
\end{equation}
The actuation depends implicitly on the ratio  $c_1/c_0$, reflecting the relative influence of bulk and hinge elasticity in the model.

In Fig.\;\ref{fig:fig3}(b), we plot $(\lambda, 2\xi(\lambda))$ for $c_1/c_0 = 0,10^{-1}, 10^{-2},$ and $10^{-3}$. The bulk elastic term dominates at small values of stretch, since the plot is essentially independent of $c_1/c_0$ in this regime.  As the stretch increases, hinge elasticity becomes more pronounced; the curve shows a particularly strong dependence on $c_1/c_0$ in the large stretch regime ($\lambda >1.2)$. We note, generally, that the slit actuation decreases monotonically as a function of increasing $c_1/c_0$, and hinge elasticity cannot be ignored at large values of stretch for this sample.  Also, the ratio $c_1/c_0 = 10^{-2}$ provides a good fit to our experimental data.  We use $c_1/c_0=10^{-2}$ from here on when comparing the deformed rotating-squares sample to analogous simulations.

\subsection{Heterogeneous loading}\label{ssec:ss44}
\begin{figure}[b!]
\centering
\includegraphics[scale=1.0]{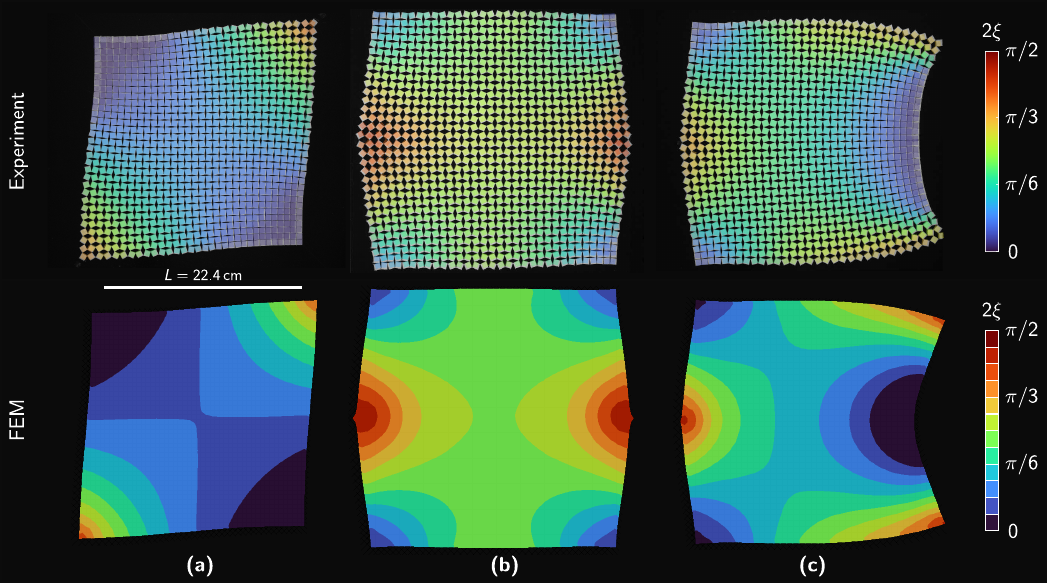}
\caption{Experiments and simulations of rotating-squares sample. Three boundary conditions are applied: (a) pulling along diagonal line; (b) pulling along centerline; (c) pulling a center point with opposite corner points fixed. $L$ is the undeformed sample length. Color maps show the slit actuation angle $\xi$, extracted from the experiment per \ref{ssec:Experimental}.}
\label{fig:fig4}
\end{figure}

\begin{figure}[t!]
\centering
\includegraphics[scale=1.0]{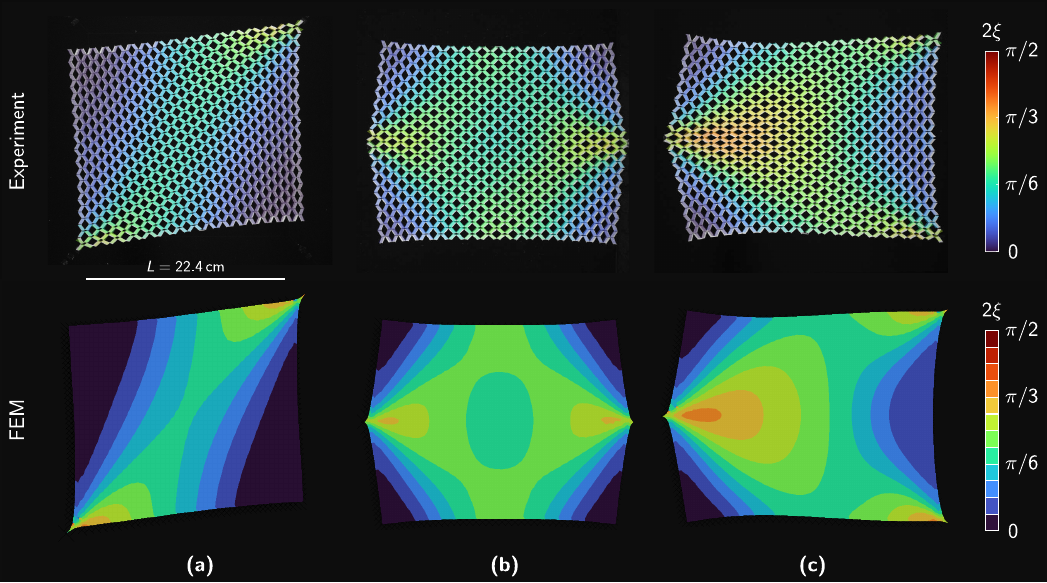}
\caption{Experiments and simulations of the hyperbolic sample. Three boundary conditions are applied: (a) pulling along diagonal line; (b) pulling along centerline; (c) pulling a center point with opposite corner points fixed.  $L$ is the undeformed sample length. Color maps show the slit actuation angle $\xi$, extracted from the experiment per \ref{ssec:Experimental}.}
\label{fig:fig5}
\end{figure}

We turn now to a comparison between FEM simulations of the model and experiments under complex loading conditions. We perform the FEM simulations in Abaqus/Standard using the UEL subroutine developed in Section \ref{ssec:FEM}:   a 2D square domain non-dimensionalized to have unit length is meshed by 1600 ($8$ node quadratic quadrilateral) user elements and is subjected to various displacement boundary conditions. The model has three moduli $c_0$, $c_1$, $c_2$ which need to be supplied for the two samples, along with the cell parameters $\alpha$ and $\beta$ from Fig.\;\ref{fig:LaserCut}.  We normalize the elasticity by  $c_0$; this choice influences the overall magnitude of the stresses, but not the predicted deformation. {\color{black}As a side note, the need to fit  $c_0$ will present itself in Section~\ref{ssec:fd}.}  Lastly, we fit the ratios $c_1/c_0$ and $c_2/c_0$ so as to accurately capture the response of a broad range of experiments. For simplicity, the fit is done by trial and error for a variety of boundary conditions. Each simulation typically takes less than  minute on a standard laptop, so this approach is not tedious or difficult. In the following comparison, we use 
\begin{equation}
\begin{aligned}\label{eq:FitModuli}
&\text{(Rotating-Squares:)} && c_1/c_0 =  10^{-2}, && c_2/c_0 = 5 \times 10^{-5}, \\
&(\text{Hyperbolic Pattern:}) &&  c_1/c_0 = 3 \times 10^{-2},  &&  c_2/c_0 = 10^{-4}.
\end{aligned}
\end{equation}
One should not extrapolate these values to other samples. We expect $c_1/c_0$ and $c_2/c_0$ to vary for patterns fabricated with a different number of cells or hinge-panel dimensions, even if the unit cell geometry and underlying material are otherwise the same.  As discussed in Section \ref{ssec:Model1}, basic physical reasoning suggests the scalings $c_1/c_0 \sim \delta^2/\ell^2$ and $c_2/c_0\sim\ell^2$ \cite{coulais2018characteristic,czajkowski2022conformal,zheng2022continuum} (up to a log factor \cite{plucinskyetal}).

With Figs.\;\ref{fig:fig4}-\ref{fig:fig5}, we demonstrate numerical simulations that accurately capture the heterogeneous engineering scale response of the two kirigami patterns under complex boundary conditions. We consider, in particular, three boundary conditions: pulling along diagonal line (Figs.\;\ref{fig:fig4}-\ref{fig:fig5}(a)), pulling along centerline (Figs.\;\ref{fig:fig4}-\ref{fig:fig5}(b)), and pulling a center point with the opposite corner points fixed (Figs.\;\ref{fig:fig4}-\ref{fig:fig5}(c)).  Each such loading leads to a soft response in the two experimental patterns, far from any pure mechanism.

Some general features emerge from the experiments. In the rotating-squares sample, slit actuation quickly decays in arcs around the loading points.  In the hyperbolic sample,  slit actuation instead radiates from the loaded boundary, yielding large bands of actuation within the sample's bulk. This ``decay versus persistence" in actuation affirms the link to Poisson's ratio derived in Ref.~\cite{zheng2022continuum} and discussed in Section \ref{ssec:StressSec} and Section \ref{ssec:ss41}. Our simulations capture these general features, as well as finer details. 

\begin{table}[h!]
\begin{center}
\begin{tabular}{||c c c c||} 
 \hline  
Boundary displacement comparison &  \Big| RMS/$\ell$ &  Max/$\ell$  &  Mean/$\ell$   \\ [.5ex] 
 \hline\hline
 Rotating-squares, Fig.\;\ref{fig:fig4}(a)  & 0.16 & 0.33 & 0.15 \\ 
 \hline
  Rotating-squares, Fig.\;\ref{fig:fig4}(b) & 0.13 & 0.28 & 0.12 \\
 \hline
   Rotating-squares, Fig.\;\ref{fig:fig4}(c) & 0.27 & 0.45  & 0.26 \\
 \hline
 Hyperbolic, Fig.\;\ref{fig:fig5}(a) & 0.24 & 0.36 & 0.22 \\
 \hline
 Hyperbolic, Fig.\;\ref{fig:fig5}(b) & 0.11 & 0.20 & 0.10 \\ 
 \hline
 Hyperbolic, Fig.\;\ref{fig:fig5}(c) & 0.25 & 0.38 & 0.23 \\ [1ex] 
 \hline
\end{tabular}
\caption{Comparison of the normed difference of  boundary displacements between experimental and simulated samples. Three metrics of this boundary value comparison are shown: the root mean squared deviation, the max, and the mean. Each is normalized by the unit cell length $\ell$ {\color{black}from Fig.\;\ref{fig:LaserCut}, which is $1.4$ cm for all experimental samples.}}
\label{tab:Table1}
\end{center}
\end{table}

With the moduli parameters of the model given by Eq.\;(\ref{eq:FitModuli}), we carry out simulations by matching the boundary conditions of the three experiments for each sample. (These boundary conditions include the displacements where the sample is loaded, as well as the natural traction-free and $\xi$ boundary conditions in Eq.\;(\ref{eq:strongForm2})). The simulations are shown on the bottom row of Fig.\;\ref{fig:fig4} and Fig.\;\ref{fig:fig5}, and each recovers the pattern's engineering scale response on the top row.  As the figures highlight, the solved-for slit actuation agrees qualitatively with that of the experiments. Table \ref{tab:Table1}  also compares various metrics of the normed difference in boundary displacement between the experimental and simulation results. The agreement is  quantitatively excellent. Each pattern is subject to roughly $10 - 30\%$ engineering strain, yet the boundary discrepancies are but a fraction of the length of a unit cell. 

\subsection{Force--displacement curves}\label{ssec:fd}

{\color{black} As a final demonstration of the capabilities of our model, we compare an experimental force--displacement curve to analogous simulations. We focus on the rotating squares architecture and the center-pulling loading condition for the demonstration. The experimental sample, introduced in Fig.\;\ref{fig:LaserCut}(a), is loaded along its centerline using a string and weights, as illustrated in Fig.\;\ref{fig:forcedispl}(a). For each value of applied force, the total stretch of the specimen along the same centerline is recorded via image processing. Additional details on the experimental setup are reported in ~\ref{ssec:Experimental}.
\begin{figure}[htb!]
\centering
\includegraphics[scale=1.0]{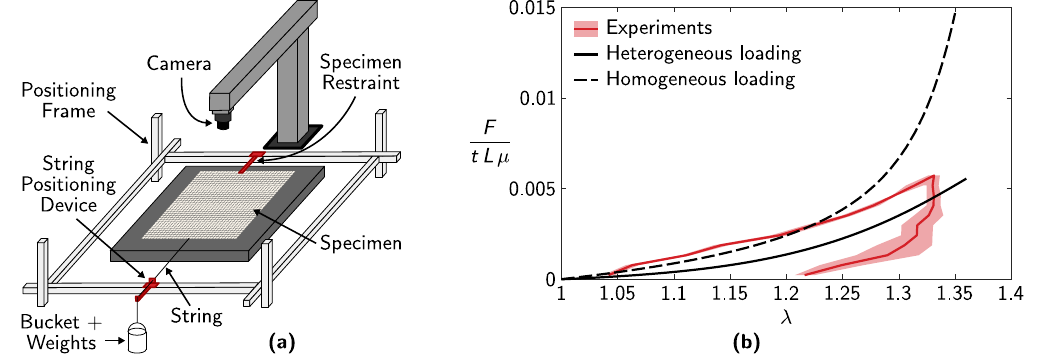}
\caption{\color{black}(a) Schematic of our experimental setup for force--displacement measurement. The specimen rests on a smooth surface and is restrained using fixtures connected to an outer frame. Forces are applied to the specimen (here, at a point where the centerline meets the boundary) through strings and weights. (b) Normalized force as a function of the centerline stretch $\lambda$. The experimental data is red, with the continuous line giving the average of three measurements and the shaded area showing the standard deviation. The solid black curve is the theoretical curve found by simulating the effective model. For comparison, the dashed curve shows the theoretical response to uniform loading, using the same fitting parameters.}
\label{fig:forcedispl}
\end{figure}

The loading and unloading experimental curves are plotted in red in Fig.\;\ref{fig:forcedispl}(b), where the solid line indicates the average and the shaded area encompasses the standard deviation from three tests. In the plot, the force is non-dimensionalized by the thickness of the sample $t = 0.15\,$cm, its width $L = 22.4\,$cm and the shear modulus of the material $\mu = 0.38\,$MPa (obtained from experimental data on the behavior of natural rubber gum in Ref.~\cite{celli2018shape}). Friction clearly plays a  role in the experiment. While the sample remains elastic throughout the entire loading and unloading process, friction causes stretches to be smaller than in a frictionless case during loading, and larger during unloading. Despite this behavior, it is still illuminating to compare this experiment to simulations of our effective (and frictionless) model. }

{\color{black} The black curve in Fig.\;\ref{fig:forcedispl}(b) plots the simulated force--displacement curve for the heterogeneous loading shown in the bottom pane of Fig.\;\ref{fig:fig4}(b) (the center-pulling case). The simulation is carried out using the rotating squares parameters $\alpha, \beta, c_1/c_0, c_2/c_0$ fitted previously. The modulus $c_0$, which has yet to be prescribed, acts as an effective shear modulus that sets the overall magnitude of the force but does not affect the shape of the curve. We choose it as $c_0 = \mu/3.5$ so that the simulated curve is approximately in the middle of the loading and unloading experimental curve. This prescription is an attempt to capture the frictionless behavior of the sample, which should fall somewhere within the width of the hysteresis loop. We note that the simulated sample is displacement-controlled rather than force-controlled (the former is easier to implement in our Abaqus UEL). Its forces are computed using a version of Castigliano's method described in \ref{sec:Castig}. We expect an analogous force-controlled simulation to produce the same force--displacement curve. 

The experimental and simulated curves exhibit the same trends and are quantitatively consistent. Both are non-linear and convex, and the modulus  used in the model ($c_0 = \mu/3.5$) is physically reasonable. In the simulation, the forces emerge from stresses due to the  non-linear bulk elastic term at a given applied displacement. This term  attempts to relax some of the actuation $\xi$ of  the linear hinge bending term. However, there is geometric frustration in this process; the effective deformation and slit actuation must approximate a local mechanism  via Eq.\;(\ref{eq:softMode}). This frustration increases as the boundary displacement increases, leading to the convex nature of the curve. That the experimental curve displays a similar convex profile is another demonstration of the quality of our generalized continuum model. 

Adding to the discussion, the dashed black line in Fig.\;\ref{fig:forcedispl}(b) plots the analytical force--displacement curve for the uniform stretch case described in Section \ref{ssec:ss42} using the same parameters as the simulation. Compared to the center-pulling case, the curve has a larger value of force at each given stretch, which  is expected since much more of the boundary is loaded at the same value of stretch.   The curve also increases dramatically at high values of stretch, where the entire pattern  has begun to exhaust its soft mechanistic response.  At this stage, the hinges must stretch as well as bend, which fully engages the  bulk elastic term in the model. Interestingly, the same level of actuation in the center pulling case ---  actuation warranting significant hinge stretching --- only occurs in localized regions  near the loading location. This feature appears to delay the inevitable sharp change in the overall force response.       }

\section{Conclusions}\label{sec:s5}
This paper modeled  the class of planar, rhombi-slit kirigami metamaterials as generalized elastic continua. We first described how to obtain the cell-averaged response of the pattern's mechanism deformations. Then we formulated an elastic energy that drove the pattern's effective deformation and slit actuation towards this cell-averaged response locally, and treated the slit actuation as an auxiliary field variable to account for additional physically relevant  sources of elasticity.  Through implementation in Abaqus, we demonstrated a model capable of predicting the response of kirigami metamaterials across designs and loading conditions.

Although we only discussed a specific energy formula for a specific family of kirigami, our method can be generalized to a wide range of kirigami metamaterials. For instance, by modifying the shape tensor in Eq.\;(\ref{eq:Axi}), we can model planar kirigami with  parallelogram slits \cite{zheng2022continuum}, and presumably any other periodic and planar kirigami with periodic mechanisms  to be discovered. We can also account for spatial variations of unit cell design, as in Ref.~\cite{celli2018shape}, by allowing the design parameters in the model to vary. Finally, while our constitutive choices in Eqs.\;(\ref{eq:strainEnergy}-\ref{eq:hookean}) were simple by design, they can be easily updated to enrich the model to account for more nuanced features of the kirigami's response than those discussed here. 

%It is worth mentioning that our mechanical model is able to predict the force--displacement curves once the value of $c_0$ in Eq.\;(\ref{eq:strainEnergy}) is assigned. We did not discuss applied forces in this paper due to the limitations (mainly friction-related) of our current experimental setup.

Taking a broader view, our results suggest that modeling mechanical metamaterials as generalized elastic continua is a powerful approach to understanding their nonlinear response --- one potentially capable of efficiently navigating the design space of these materials, while remaining predictive under a wide range of loads. By focusing on rhombi-slit kirigami,  we took  a purposefully concrete and simple approach to modeling.  We hope this choice makes our work widely accessible and  paves the way for broad generalization going forward.

\section*{Acknowledgements}

\noindent Y.Z.\ and P.P.\ acknowledge support through P.P.’s startup package at the University of Southern California. I.T.\ acknowledges support from the National Science Foundation (DMS-CAREER-2145225). P.C. {\color{black}and I.N.}\ acknowledge support from the National Science Foundation (CMMI-2045191).

\section*{Author contributions}
\noindent Y.Z.:  Conceptualization, Formal Analysis, Investigation, Methodology, Software, Writing - original draft. {\color{black}I.N.: Investigation.} I.T.: Conceptualization, Methodology, Writing - review and editing. P.C.: Conceptualization, Visualization, Investigation, Writing - review and editing. P.P.: Conceptualization, Methodology, Supervision, Writing - original draft

\section*{Data availability}

\noindent The Abaqus UEL codes are available on GitHub 
 at:\\ \url{https://github.com/yzheng29/kirigami-metamaterial\_UEL}

\appendix

\section{Explicit tangent formulas for Abaqus FEM implementation}\label{sec:Appendix}

 \renewcommand{\theequation}{A \arabic{equation}}

Here, we develop explicit formulas for the tangents in Eq.\;(\ref{eq:stiffnessMeasures}) to complete the description of the FEM formulation for Abaqus implementation. The formulas reference functions introduced in Sections \ref{sec:Model} and \ref{sec:EquilFEA} which we do not repeat here. The formulas also make use of shape functions defined on the current configuration  $\tilde{N}^A \colon \mathbf{y}_{\text{eff}}(\Omega) \rightarrow \mathbb{R}$ via $\tilde{N}^A \circ \mathbf{y}_{\text{eff}}( \mathbf{x}) = N^A(\mathbf{x})$, for which  the gradients transform as 
\begin{equation}
\begin{aligned}
\nabla \tilde{N}^A \circ \mathbf{y}_{\text{eff}}(\mathbf{x}) = \big(\tilde{\mathbf{F}}(\mathbf{x}, \mathbf{u})\big)^{-T} \nabla N^A(\mathbf{x}).
\end{aligned}
\end{equation}
This transformation is a convenient way of bringing out certain symmetries in the parts of these tangents associated to the bulk elastic term $W_0$. (Chapter 8.4 \cite{bower2009applied} has a detailed exposition on hyperelastic tangent formulas, where this transformation is employed.) In brief, we find that 
\begin{equation}
    \begin{aligned}\label{eq:kStiffInt}
    &\big[ \mathbf{K}_{\text{disp}}^{AB}(\mathbf{u}, \boldsymbol{\xi}) \big]_{ab} = \int_{\Omega}\Big\{  \Big[\mathbb{C}_{\text{disp}}\big(\tilde{\mathbf{F}}(\mathbf{x}, \mathbf{u})  \mathbf{A}^{-1}(\tilde{\theta}(\mathbf{x}, \boldsymbol{\xi}))\big)\Big]_{abcd} \big[\nabla \tilde{N}^{A} \circ \mathbf{y}_{\text{eff}}(\mathbf{x})\big]_c \big[\nabla \tilde{N}^{B} \circ \mathbf{y}_{\text{eff}}(\mathbf{x})\big]_d \Big\}  dA, \\
    &\big[ \mathbf{k}_{\text{disp}}^{AB}(\mathbf{u},\boldsymbol{\xi}) \big]_{a} = \big[ \mathbf{k}_{\text{act}}^{BA}(\mathbf{u}, \boldsymbol{\xi}) \big]_{a} =  \int_{\Omega} \Big\{  \big[ \mathbf{C}_{\text{mix}}\big(\tilde{\mathbf{F}}(\mathbf{x}, \mathbf{u}) , \tilde{\theta}(\mathbf{x}, \boldsymbol{\xi}))\big)\big]_{ab} \big[\nabla \tilde{N}^B \circ \mathbf{y}_{\text{eff}}(\mathbf{x}) \big]_b N^A(\mathbf{x}) \Big\} dA, \\ 
    &k_{\text{act}}^{AB}(\mathbf{u}, \boldsymbol{\xi}) = \int_{\Omega}\Big\{ \frac{\partial}{\partial \theta}\big[ f_{\text{act}}\big(\tilde{\mathbf{F}}(\mathbf{x}, \mathbf{u}) , \tilde{\theta}(\mathbf{x}, \boldsymbol{\xi})\big)\big] N^A(\mathbf{x}) N^B(\mathbf{x}) +  2c_2 \nabla N^{A}(\mathbf{x}) \cdot \nabla N^B(\mathbf{x}) \Big\} dA
    \end{aligned}
\end{equation}
in 2D index notation  with repeated indices summed. The moduli in these formula are associated to partial derivatives of $W(\mathbf{F}, \theta, \mathbf{p})$.  Structurally, they are of the form
\begin{equation}
    \begin{aligned}\label{eq:GetModuli}
    &\big[\mathbb{C}_{\text{disp}}(\mathbf{G}) \big]_{abcd} 
    = \Big[ \frac{\partial \boldsymbol{\tau}_0(\mathbf{G})}{\partial \mathbf{G}}\Big]_{bdal} \big[\mathbf{G}\big]_{cl} -   \big[\boldsymbol{\tau}_0(\mathbf{G})\big]_{bc} \big[\mathbf{I}\big]_{ad}, \\
    &\mathbf{C}_{\text{mix}}(\mathbf{F}, \theta) := \widetilde{\mathbf{C}}_{\text{mix}}\big(\mathbf{F}\mathbf{A}^{-1}(\theta), \mathbf{A}'(\theta) \mathbf{A}^{-1}(\theta)\big), \quad  \widetilde{\mathbf{C}}_{\text{mix}}\big(\mathbf{G}, \mathbf{H})  =  -\frac{\partial \boldsymbol{\tau}_0(\mathbf{G})}{\partial \mathbf{G}} \colon  \big(\mathbf{G}\mathbf{H}\big), \\
    &\frac{\partial}{\partial \theta} f_{\text{act}}(\mathbf{F}, \theta) = -\frac{d}{d\theta} \Big[ \mathbf{P}_0(\mathbf{G}(\theta)) \colon \big(\mathbf{G}(\theta) \mathbf{H}(\theta)\big) \Big] + 2c_1 =: c_{\text{act}}\big(\mathbf{G}(\theta), \mathbf{H}(\theta),\mathbf{H}'(\theta)\big), 
    \end{aligned}
\end{equation}
where the arguments of $c_{\text{act}}(\cdot)$ are defined as $\mathbf{G}(\theta) := \mathbf{F} \mathbf{A}^{-1}(\theta)$ and $\mathbf{H}(\theta) := \mathbf{A}'(\theta) \mathbf{A}^{-1}(\theta)$. Each moduli can also be expressed in terms of elementary functions as  
\begin{equation}
    \begin{aligned}
    &\big[\mathbb{C}_{\text{disp}}(\mathbf{G}) \big]_{abcd} 
    = \frac{2c_0}{J_{\mathbf{G}}} \Big\{ [\mathbf{I}]_{ba} [\mathbf{G}\mathbf{G}^T]_{dc} - \big[ \mathbf{G} \mathbf{G}^T \big]_{bd} \big[ \mathbf{I} \big]_{ac}  - \big[\mathbf{I}\big]_{bd} \big[ \mathbf{G} \mathbf{G}^T\big]_{ac}  + (2 J_{\mathbf{G}}^3 - J_{\mathbf{G}}^2) \big[\mathbf{I}\big]_{bd} \big[\mathbf{I}\big]_{ac} \Big\}  \\
    &\qquad \qquad \qquad \qquad   + \frac{2c_0}{J_{\mathbf{G}}}\Big\{  -(J_{\mathbf{G}}^3 -J_{\mathbf{G}}^2)\big[\mathbf{I}\big]_{bc}\big[\mathbf{I}\big]_{ad} +\frac{1}{2}|\mathbf{G}|^2\big(\big[\mathbf{I}\big]_{bd}\big[ \mathbf{I} \big]_{ac} + \big[\mathbf{I}\big]_{bc}\big[\mathbf{I}\big]_{ad} \big)\Big\}, \\
     &\widetilde{\mathbf{C}}_{\text{mix}}(\mathbf{G}, \mathbf{H}) = -\frac{2 c_0}{J_{\mathbf{G}}} \Big(2 \sym \big( \mathbf{G} \mathbf{H}\mathbf{G}^T \big) - \Tr(\mathbf{G} \mathbf{H} \mathbf{G}^T) \mathbf{I}  +  \Tr(\mathbf{H}) \big\{(\frac{1}{2} |\mathbf{G}|^2 + 2 J_{\mathbf{G}}^3 - J_{\mathbf{G}}^2) \mathbf{I} - \mathbf{G} \mathbf{G}^T \big\} \Big),  \\
&c_{\text{act}}(\mathbf{G}, \mathbf{H}, \mathbf{M}) = \frac{2 c_0}{J_{\mathbf{G}}} \Big( |\mathbf{G} \mathbf{H}|^2 + \Tr \big( \mathbf{G} ( \mathbf{H}^2 - \mathbf{M}) \mathbf{G}^T\big) - 2  \Tr( \mathbf{G} \mathbf{H} \mathbf{G}^T) \Tr( \mathbf{H})  \Big)    \\
&\qquad \qquad \qquad \qquad + \frac{2 c_0}{J_{\mathbf{G}}} \Big(\big\{2 J_{\mathbf{G}}^3 - J_{\mathbf{G}}^2 + \tfrac{1}{2}|\mathbf{G}|^2 \big\} \big(\Tr(\mathbf{H})\big)^2 + \big\{ \tfrac{1}{2} |\mathbf{G}|^2 - J_{\mathbf{G}}^3 + J_{\mathbf{G}}^2\big\} \Tr(\mathbf{M}) \Big) + 2c_1,
    \end{aligned}
\end{equation}
where $\sym(\cdot )$ denotes the symmetric part of a square matrix.

\section{Materials and methods}\label{ssec:Experimental}

 \renewcommand{\theequation}{B \arabic{equation}}

We cut the  kirigami specimens out of 1.5 mm-thick natural rubber sheets (McMaster-Carr 8633K71) using an 80 Watt Epilog Fusion Pro 32 laser cutter. During the fabrication, the laser cutter is focused on the bottom face of the rubber sheet to avoid burning the specimens and to produce clean cuts. The specimens are painted with white primer paint  to create high contrast with a black background, which  facilitates image processing. Images of the samples are recorded by means of a FLIR 5-megapixel 35 fps camera with Edmund Optics lenses.

Our method  for characterizing  the slit actuation in the deformed experimental samples is similar to our previous work \cite{zheng2022continuum} and to that of others \cite{deng2020characterization}. We obtain quantitative information on the deformation through digital image processing in MATLAB. After converting the images to binary (using the \verb+imbinarize+ function), we obtain the centroid $\mathbf{c}^{(i,j)}$, semi-major axis $\mathbf{a}^{(i,j)}$, and semi-minor axis $\mathbf{b}^{(i,j)}$ of the central slit of each $(i,j)$-cell using the \verb+regionprops+ function. Then, using $d = \ell/20$ from Section \ref{ssec:ss41} and $\xi_0$ from Fig.\;\ref{fig:LaserCut}(c-d), we calculate $\xi^{(i,j)}$ from $\tan (\xi^{(i,j)} + \xi_{0}) = (|\mathbf{a}^{(i,j)}|+d/2)/(|\mathbf{b}^{(i,j)}|+d/2)$ and take $\gamma^{(i,j)}$ as the inclination of the major axis with respect to the horizontal. Since we know the dimensions of the panels in the sample, we overlay a $\xi$ color map of each idealized deformed unit cell (with angles ($\xi^{(i,j)}$,$\gamma^{(i,j)}$)) onto the experimental pattern, centered at the slit centroid $\mathbf{c}^{(i,j)}$.

{\color{black} The force--displacement data is obtained with the setup illustrated in Fig.~\ref{fig:forcedispl}(a). The specimen rests on a smooth surface (Slippery UHMW Polyethylene). For anchoring and positioning purposes, we build a frame of T-slotted rails and use a 3D-printed hook to pin down the specimen onto the surface. At the desired loading location, we tie a nylon string to the specimen. The string then passes through a 3D-printed positioning device tied to the frame, and is connected to a bucket. We add known weights to the bucket and calculate the force that the weights impart onto the specimen. These experiments are force-controlled: we apply a weight and measure the specimen's stretch along a desired direction. The stretch is measured with the same camera mentioned previously in this section. The deformed length of the specimen is measured through an automated pixel counting process in MATLAB.}

 \section{Castigliano's method to obtain forces}\label{sec:Castig}

 \renewcommand{\theequation}{C \arabic{equation}}
 
{\color{black} The theoretical forces in Fig.\;\ref{fig:forcedispl} are 
 calculated by differentiating the equilibrium energy with respect to the overall displacement, using our generalized elastic continuum model. This is essentially Castigliano's method from structural mechanics. Here, we give a brief justification of this approach.

We assume that the effective reference domain of the kirigami pattern is $\Omega = (-L/2,L/2)^2$ and focus on the center-pulling case, where the pattern's effective deformation $\mathbf{y}_{\text{eff}}(\mathbf{x})$ and slit actuation $\xi(\mathbf{x})$ solve the equilibrium equations in Eq.\;(\ref{eq:strongForm})  subject to the boundary conditions 
\begin{equation}
\begin{aligned}\label{eq:BCsCenterPull}
\begin{cases}
\mathbf{y}_{\text{eff}}(\mathbf{x}) = \mathbf{x} \pm  \frac{q}{2} 
 \mathbf{e}_1 & \text{for }\mathbf{x} \cdot \mathbf{e}_1 = \pm L/2,\  \mathbf{x} \cdot \mathbf{e}_2 \in (-\epsilon, \epsilon)  \\
 \mathbf{P}_{\mathbf{y}_{\text{eff}},\xi}(\mathbf{x}) \mathbf{n}_\text{R}(\mathbf{x}) = \mathbf{0}\quad &\text{on the rest of the boundary $\partial\Omega$}  \\
 \nabla \xi(\mathbf{x})\cdot \mathbf{n}_{\text{R}}(\mathbf{x}) = 0 & \text{on all of $\partial \Omega$}.
 \end{cases}
\end{aligned}
\end{equation}
The parameter $\epsilon >0$ idealizes the width where the sample is gripped and displaced on center-pulling. The parameter $q\geq 0$  is the overall displacement of the horizontal centerline as the sample is monotonically loaded. It starts at the value $q = 0$ and increases monotonically to achieve a maximum stretch $\lambda = 1 + q/L \approx 1.35$ in both the simulation and experiment in Fig.\;\ref{fig:forcedispl}. Of course, $\mathbf{y}_{\text{eff}}$ and $\xi$ depend on $q$.

As in Castigliano's method, we differentiate the equilibrium value of the generalized elastic continuum energy
\begin{equation}
    \mathcal{U}(q) := t \int_{\Omega} W(\nabla \mathbf{y}_{\text{eff}}(\mathbf{x}), \xi(\mathbf{x}), \nabla \xi(\mathbf{x}) ) dA
    \end{equation}
with respect to the overall displacement $q$. Note $t>0$ is the thickness of the sample. Passing the derivative under the integral sign and using the definitions from Sections \ref{sec:Model} and \ref{sec:EquilFEA}, there follows
\begin{equation}
\frac{d \mathcal{U}}{d q}  = t \int_{\Omega} \Big\{ \mathbf{P}_{\mathbf{y}_{\text{eff}}, \xi} (\mathbf{x}) \colon \nabla\Big( \frac{\partial \mathbf{y}_{\text{eff}}(\mathbf{x})}{\partial q}  \Big) + f_{\text{act}}( \nabla \mathbf{y}_{\text{eff}}(\mathbf{x}), \xi(\mathbf{x})) \frac{\partial \xi(\mathbf{x})}{\partial q}  + 2 c_2 \nabla \xi(\mathbf{x}) \cdot \nabla \big(\frac{\partial \xi(\mathbf{x})}{\partial q} \big) \Big\}  dA.
\end{equation}
By the divergence theorem,
\begin{equation}
\begin{aligned}\label{eq:almostDone}
\frac{d \mathcal{U}}{d q}  &=t \int_{\Omega} \Big\{  -\Big( \nabla \cdot \mathbf{P}^T_{\mathbf{y}_{\text{eff}}, \xi}(\mathbf{x}) \Big) \cdot \frac{\partial  \mathbf{y}_{\text{eff}}(\mathbf{x})}{\partial q} + \Big( f_{\text{act}}( \nabla \mathbf{y}_{\text{eff}}(\mathbf{x}), \xi(\mathbf{x})) - 2 c_2 \nabla^2 \xi(\mathbf{x}) \Big) \frac{\partial \xi(\mathbf{x})}{\partial q} \Big\} dA \\
&\quad + t\int_{\partial \Omega} \Big\{\Big(  \mathbf{P}_{\mathbf{y}_{\text{eff}}, \xi}(\mathbf{x}) \mathbf{n}_{\text{R}}(\mathbf{x})\Big) \cdot \frac{\partial  \mathbf{y}_{\text{eff}}(\mathbf{x})}{\partial q} +\Big( 2 c_2 \nabla \xi(\mathbf{x}) \cdot \mathbf{n}_{\text{R}}(\mathbf{x}) \Big)
 \frac{\partial \xi(\mathbf{x})}{\partial q } \Big\} d \Gamma. 
\end{aligned}
\end{equation}
The equilibrium equations in Eq.\;(\ref{eq:strongForm}) imply that the first integral above vanishes. Concerning the second, let $\mathbf{x}^{\pm}(s) = \pm (L/2) \mathbf{e}_1 + s \mathbf{e}_2$ parameterize the boundaries of the applied  displacement. By the boundary conditions in Eq.\;(\ref{eq:BCsCenterPull}), 
\begin{equation}
\begin{aligned}
\frac{d \mathcal{U}}{dq} &=  t \int_{-\epsilon}^{\epsilon} \Big\{ \mathbf{P}_{\mathbf{y}_{\text{eff}},\xi}\big( \mathbf{x}^+(s)\big) \mathbf{e}_1  \cdot \frac{\partial}{\partial q} \Big( \mathbf{x}^+(s) + \frac{q}{2} \mathbf{e}_1\Big) - \mathbf{P}_{\mathbf{y}_{\text{eff}},\xi}\big( \mathbf{x}^-(s)\big) \mathbf{e}_1  \cdot \frac{\partial}{\partial q} \Big( \mathbf{x}^-(s) - \frac{q}{2} \mathbf{e}_1\Big) \Big\}  ds \\
&= \underbrace{\frac{t}{2} \int_{-\epsilon}^{\epsilon} \mathbf{e}_1 \cdot \mathbf{P}_{\mathbf{y}_{\text{eff}},\xi}\big( \mathbf{x}^+(s)\big) \mathbf{e}_1 ds}_{:= \frac{1}{2} f^+}  + \underbrace{\frac{t}{2} \int_{-\epsilon}^{\epsilon} \mathbf{e}_1 \cdot \mathbf{P}_{\mathbf{y}_{\text{eff}},\xi}\big( \mathbf{x}^-(s)\big) \mathbf{e}_1 ds}_{:=\frac{1}{2}f^{-}} .
\end{aligned}
\end{equation}
The terms $f^+$ and $f^-$ give the tensile force applied to the right and left boundary, respectively. As these are equal by force balance, i.e., $f^+ = f^- = F$,  we conclude that
\begin{equation}
\begin{aligned}\label{eq:Castig}
\frac{d \mathcal{U}}{dq}  = F.
\end{aligned}
\end{equation}

In summary, the force in the center-pulling setup can be found by differentiating the equilibrium energy with respect to the overall displacement. Eq.\;(\ref{eq:Castig}) also holds for the homogeneous setup in Fig.\;\ref{fig:forcedispl}. The justification of this result is similar and is left to the reader.}

\bibliographystyle{unsrt}
% \bibliography{bib}

\end{document}